\documentclass[a4paper,10pt]{article}
\usepackage{xspace}
\usepackage{amsbsy} 
\usepackage{graphicx}

\usepackage[left=2.5cm,right=2.5cm,top=2cm,bottom=2cm]{geometry}

\usepackage{subfigure}
\usepackage{color}
\usepackage{setspace}
\usepackage{multirow} 
\usepackage{multicol}
\usepackage[version=3]{mhchem}

\usepackage{fancyhdr}
\fancypagestyle{plain}{ %
\fancyhf{} 
\lhead{\textit{Structural relationship; \ce{Mg4Zn7},\ce{MgZn2}}}
\rhead{J.M.Rosalie, A.Singh, H.Somekawa \& T.Mukai}
\lfoot{\textit{Corrected proof}}
\rfoot{\thepage}
}

\newcommand{\betap}{\ensuremath{{\beta_1^\prime}}\xspace}
\newcommand{\betapp}{\ensuremath{{\beta_2^\prime}}\xspace}
\newcommand{\iphasecomp}{\ensuremath{\mathrm{Mg_3ZnY_6}}\xspace}
\newcommand{\micron}{\ensuremath{\mathrm{\mu m}}\xspace}
\newcommand{\degree}{\ensuremath{\mathrm{^o}}\xspace} 
\newcommand{\celsius}{\ensuremath{\mathrm{^oC}}\xspace} 

\newcommand{\fig}[2][6cm]{\resizebox{#1}{!}{\includegraphics{#2}}} 

\begin{document}

\title{Structural relationships among monoclinic and hexagonal phases and transition structures in Mg-Zn-Y alloys} 

\author{Julian~M~Rosalie$\mathrm{^{a}}$ \and Hidetoshi Somekawa$\mathrm{^{a}}$ \and Alok Singh$\mathrm{^{a}}$,
 \and Toshiji Mukai$\mathrm{^{a}}$}
\date{}

\maketitle
\noindent
$\mathrm{^{a}}$Lightweight Alloys Group, Structural Metals Center, 
National Institute for Materials Science (NIMS), Sengen 1-2-1, Tsukuba,
Ibaraki, 305-0047, Japan.
\vspace{6pt}
\maketitle

\begin{abstract}
Isothermal ageing of plastically deformed Mg-Zn-Y alloys resulted in precipitation along 
$\{10\overline{1}2\}$ twin boundaries. 
The bulky precipitates formed had structures similar to those recently reported for the rod-like \betap precipitates, but afforded a more detailed study by high resolution TEM due to their larger size. 
The core of the precipitates often had the structure of monoclinic \ce{Mg4Zn7} phase, and had the orientation 
\( [0001]_\mathrm{Mg}\parallel[010]\); \(\{10\overline{1}0\}_\mathrm{Mg}\parallel (201)\) with either the matrix or the twin. 
On this \ce{Mg4Zn7} phase, hexagonal \ce{MgZn2} phase grew in two orientations, both with 
\([010]_\ce{Mg4Zn7} \parallel [11\overline{2}0]_\ce{MgZn2}\). 
One of these orientations formed a known orientation relationship 
\([0001]_\mathrm{Mg}\parallel [11\overline{2}0]_\ce{MgZn2}\); 
\(\{11\overline{2}0\}_\mathrm{Mg}\parallel (0001)_\ce{MgZn2}\) with the matrix. 
The part of the precipitate with the \ce{MgZn2} structure was usually in direct contact with the twin boundary. 
Both \ce{Mg4Zn7} and \ce{MgZn2} phases have layered structures that can be described with similar building blocks of icosahedrally coordinated atoms. 
The atomic positions of zinc atoms comprise the vertices of these icosahedra and form ``thick'' rhombic tiles. 
Orientation of these rhombuses remain unchanged across the interfaces between the two phases. 
Near the interface with \ce{MgZn2}, transition structures formed in \ce{Mg4Zn7} phase, with Zn:Mg atom ratio between those of \ce{Mg4Zn7} and \ce{MgZn2} phases. 
In these transition structures, the unit cell of \ce{Mg4Zn7} phase is extended along [100] or [001] by half a unit cell length by continuation of the rhombic tiling. 
Structures of these extended unit cells are proposed. 
\end{abstract}


\section{Introduction}

Magnesium alloys find application in automotive components and other areas where their strength to weight ratio  is advantageous.
The magnesium-zinc alloy system possesses the greatest precipitation hardening response of commercial Mg-based alloys \cite{ClarkMgZn1965}  
 due to the precipitation of a fine,
rod-like precipitate termed \betap\cite{ClarkMgZn1965,Sturkey1959,MimaAgeing1971}. 
These precipitates provide resistance to basal slip, being highly elongated, with aspect ratios of around 40:1, and resistant to shearing by dislocations\cite{ClarkMgZn1965}. 
In fully aged samples the precipitates also significantly restrict deformation by twinning \cite{Chun1969}. 

Despite the importance of this phase in strengthening Mg-Zn alloys, there has been on-going uncertainty regarding its structure. 
Early studies concluded that the rod-like precipitates possessed a \ce{MgZn2} \(C14\) hexagonal Laves phase structure with lattice parameters a=0.52\,nm, c = 0.85\,nm\cite{Sturkey1959,Takahashi1973}.  
This view has also been supported by more recent studies \cite{Mendis2009}.
Other investigators have proposed that the structure is instead that of a \ce{Mg4Zn7} monoclinic phase with lattice parameters
a=2.596\,nm, b=0.524\,nm, c=1.428\,nm, \(\beta =102.5\degree\) \cite{Yarmolyuk1975} and with a complex domain structure \cite{Gao2007,Singh2007}. 

Several orientation relationships (ORs) have been reported for the \ce{Mg4Zn7} phase in magnesium.  
In each case, the \( [010]_{\betap}\) and \( [0001]_{\mathrm{Mg}}\) directions lie parallel to one another, 
permitting good matching between the \(b\)-lattice parameter (0.524\,nm) and the hexagonal axis of magnesium (0.52\,nm). 
The \((01\overline{1}0)_{\mathrm{Mg}}\) planes were aligned parallel to the  
\((603)_{\betap}\) planes \cite{Gao2007,Singh2007}. 
Other possibilities were orientation relationships with 
 \((01\overline{1}0)_{\mathrm{Mg}}\) parallel to the \((200)_{\betap}\) or \((202)_{\betap}\) planes, or \( (\overline{2}110)_{\mathrm{Mg}}  \parallel (\overline{2}01)_{\betap} \) \cite{Takahashi1973, Singh2007}. 

The \betap precipitate is metastable and further ageing of Mg-Zn alloys results in the precipitation of a plate-like phase, 
termed \(\mathrm{\beta_2^{\prime}}\), 
on the basal planes of the Mg matrix.
It is accepted that this hexagonal phase does possess the \(C14\) hexagonal Laves phase structure 
(with lattice parameters a=0.52\,nm, c = 0.85\,nm \cite{Sturkey1959}) and orientation relationship \( [0001]_{\mathrm{Mg}}\parallel [0001]_{\ce{MgZn2}} \);
\( (11\overline{2}0)_{\mathrm{Mg}}\parallel (10\overline{1}0)_{\mathrm{MgZn_2}} \) 
\cite{Takahashi1973,SinghTMS2008}. 
A lath-like form with 
\( [0001]_\mathrm{Mg}\parallel [11\overline{2}0]_{\ce{MgZn2}} \); 
\( (11\overline{2}0)_\mathrm{Mg}\parallel (0001)_{\ce{MgZn2}} \) \cite{Gao2007}  
has also been reported as a minor variant.

Similar rod-shaped (\betap) and plate-shaped (\betapp) precipitates are also observed in Mg-Zn-RE/Y (RE=Rare earth) alloys.
These precipitates are thought to be identical in structure and composition to those in Mg-Zn alloys since  rare earth elements and yttrium segregate to ternary phase precipitates at the grain boundaries \cite{WeiPrec1995} and have not been detected within the \betap and \betapp precipitates \cite{Singh2007, WeiPrec1995}. 

Among the ternary phases also exists the quasi-crystalline icosahedral phase (I-phase, \iphasecomp)\cite{Luo1993,Tsai1994}. This phase exists in equilibrium with $\alpha$-Mg phase, forming a two phase field when the atomic ratio of Zn:Y is near 6, as in the alloy studied here.

The monoclinic \ce{Mg4Zn7} and Laves phase \ce{MgZn2} phases include zinc lattice sites which possess icosahedral symmetry and are thus are structurally related.  
The monoclinic \ce{Mg4Zn7} phase has been observed to co-exist with a decagonal quasi-crystalline approximant phase \cite{Yi2002} and recent work has 
shown that \ce{Mg4Zn7} and \ce{MgZn2} phases co-exist within the rod-like \betap precipitates in Mg-Zn-Y alloys \cite{Singh2009coexist}.
There appear to be structural similarities between the binary phases and the icosahedral phase and the monoclinic phase provides a model for the icosahedral structure\cite{Yang1987}. 
A proper understanding of the relationships between the monoclinic and hexagonal phases would assist in clarifying their structural relationship to the i-phase and resolve the uncertainty regarding the phase present in the rod-shaped \betap precipitates.

In the present work, extrusion followed by slight compression has been used to produce twins.  Precipitation occurred along twin boundaries upon ageing. 
Twin boundary precipitates have been identified as belonging to \ce{Mg4Zn7} and  \ce{MgZn2} phases and to be larger in scale than the \betap rods, with similar domain structures \cite{RosalieTms2009}. 
The larger size of the twin-boundary precipitates has offered an opportunity to examine the relationships between these precipitates in greater detail. 

\section{Experimental details}
A magnesium alloy containing 2.8at.\%Zn and 0.4at.\%Y was produced via direct chill casting. 
The billet was homogenised at 350\celsius, for 15\,h and extruded with extrusion ratio 12:1 at 300\celsius to produce a 12\,mm diameter rod. 
The extrusion process produced a tube texture, in which the basal planes lie parallel to the extrusion direction. 
For this texture, compression in the extrusion direction corresponds to \(c\)-axis tension of Mg and 
results in extensive \(\{10\overline{1}2\}\) twinning at stresses just above the yield point \cite{Kelley1968,BarnettTwins2007A}. 

Cylindrical compression samples were machined from the extruded rod, encapsulated in helium, solution treated (400\celsius, 1\,h) and quenched in water. 
An Instron mechanical tester was used to apply a total of 2.2\% plastic strain at a strain rate of \(1\times10^{-3}\,\mbox{s}^{-1}\). 
The level of the plastic strain was chosen so as to be sufficient to generate deformation twins while limiting slip and dislocation multiplication. 
The alloys were aged at 150\celsius in an oil bath for periods from 8\,h to 48\,h. 
Foils for TEM examination were prepared from the aged specimens, ground to \(\sim70\,\micron\)  and then thinned to perforation using a Gatan precision ion polishing system. 
A JEOL 4000EX instrument operating at 400\,kV was used for high-resolution transmission electron microscopic analysis.
Multi-slice image simulations were carried out using JEMS  software 
using the pre-loaded parameters for the the JEOL 4000EX microscope \cite{StadelmannJems1987}.

\section{Results}

\begin{figure}[htbp]
\begin{center}
\fig[9cm]{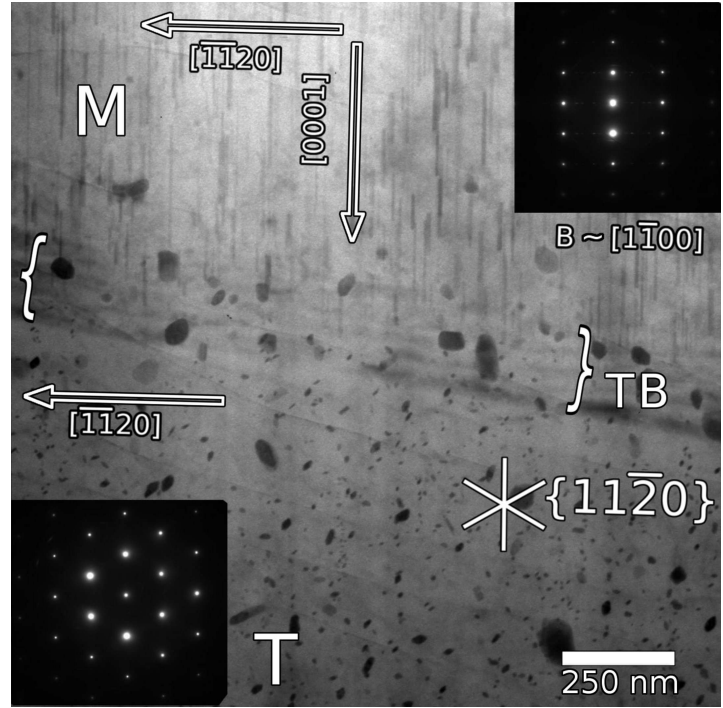}
\caption{Electron micrograph showing the microstructure of the Mg-Zn-Y alloy close to a twin boundary.
The matrix  (M, upper part of the image) is viewed along the [0001] zone while matrix (T, lower part of the image) is close to the 
\( [1\overline{1}00] \) zone. 
Rod-like \betap precipitates can be seen side-on on the matrix and in cross section in the twin. 
The twin boundary (TB) is inclined to the beam direction and is indicated by the bracketed region. 
A number of considerably larger, blocky precipitates can be seen along the twin boundary. \label{tbmicro} }
\end{center}
\end{figure}

The microstructure of the alloys consisted of rod-like \betap precipitates within both the magnesium matrix and the deformation twins, with larger, irregular precipitates present at the twin boundaries. 
Figure~\ref{tbmicro} presents a typical twin-boundary and the surrounding region.  
The matrix (M) is close to \([1\overline{1}00]_\mathrm{Mg}\) orientation, with the twin viewed along the \([0001]_\mathrm{Mg}\) zone.  
Rod-like \betap precipitates can be seen side-on in the matrix and in cross-section in the twin. 
The twin boundary \((TB)\) is inclined to the beam direction and is indicated by the bracketed region. 
A number of considerably larger, blocky precipitates can be seen along the twin boundary.
The present work is concerned exclusively with twin-boundary precipitates, whose larger size permitted a close examination of their structure.

\begin{figure}[htbp]
\begin{center}
\fig[12cm]{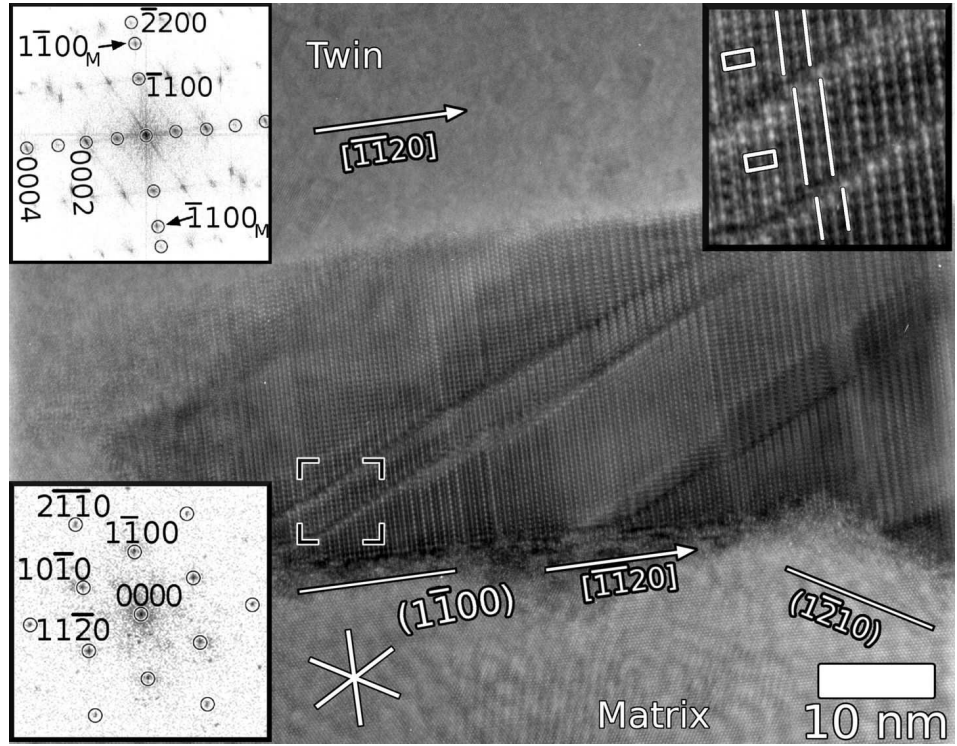}
\caption{An electron micrograph of a \ce{MgZn2} hexagonal Laves phase precipitate on a twin boundary in Mg-Zn-Y. 
The precipitate is in the lath-like orientation to the matrix with 
\( [0001]_\mathrm{Mg}\parallel [11\overline{2}0]_{\ce{MgZn2}} \)
\( (11\overline{2}0)_\mathrm{Mg}\parallel (0001)_{\ce{MgZn2}} \).
The twin boundary is indicated by a dashed line. 
\label{laves-tbp}}
\end{center}
\end{figure}

Hexagonal Laves phase precipitates were frequently observed on the twin boundaries, an example is shown in Figure~\ref{laves-tbp}. 
The matrix (lower portion) is viewed along the hexagonal axis (shown by FFT inset lower left) while the twin is 4$^o$ from its $[01\overline{1}0]$ zone. 
The $[\overline{11}20]$ direction is common to both the regions. 
The precipitate has the known orientation relationship 
$[0001]_\mathrm{Mg}\parallel [11\overline{2}0]_\ce{MgZn2}$; 
$(11\overline{2}0)_\mathrm{Mg} \parallel (0001)_\ce{MgZn2} $ 
with respect to the magnesium matrix, as shown by FFT inset upper left. 
The interface with the matrix is faceted on the $(1\overline{1}00)_{Mg}$ planes; facets on 
\( (1\overline{2}10)_{Mg} \)
planes were also observed. 
The precipitate-twin interfaces which appear planar here were curved in most instances. 
The orientation relationship between the precipitate and the twin is
$ [11\overline{2}0]_\mathrm{Mg} \parallel [0001]_\ce{MgZn2}$ ; $(0001)_\mathrm{Mg} \sim (\overline{1}100)_\ce{MgZn2}$. 
This retains the matching between $(11\overline{2}0)_\mathrm{Mg}$ and $(0001)_\ce{MgZn2}$ planes which exists between this precipitate and the matrix. 

Planar defects were observed in these \ce{MgZn2} precipitates, which were manifested in the FFT pattern like the one shown on the upper left in Figure~\ref{laves-tbp}. 
In this micrograph, the most visible defects are the diagonal lines across. 
Due to these defects, streaking occurred across the spots in the \ce{MgZn2} pattern. 
These were of two kinds. 
When the diagonal defects were excluded from FFT, streaking occurred only along 
$c^*_\ce{MgZn2}$ direction. 
This indicates stacking defects along the hexagonal axis. 
When the diagonal lines were included in the FFT, strong streaking occurred only along the 
$[\overline{1}101]$ reciprocal direction, 
perpendicular to the direction of these lines in the micrograph. 
The significance of these two kind of planar defects will be discussed in a later section. 

Precipitates composed solely of the monoclinic phase were not observed on the twin boundaries. 
However, high resolution TEM images of the  precipitates frequently showed regions of both \ce{Mg4Zn7} and \ce{MgZn2} phases within the same precipitate. 
The precipitates in which the \ce{Mg4Zn7} and hexagonal Laves phases co-existed consisted of a central region of the monoclinic phase, 
with one or two variants of the hexagonal Laves phase in contact with it and with the magnesium matrix or twin.

\begin{figure}[htbp]
\begin{center}
\subfigure[\label{hrtem1a}]{\fig[8cm]{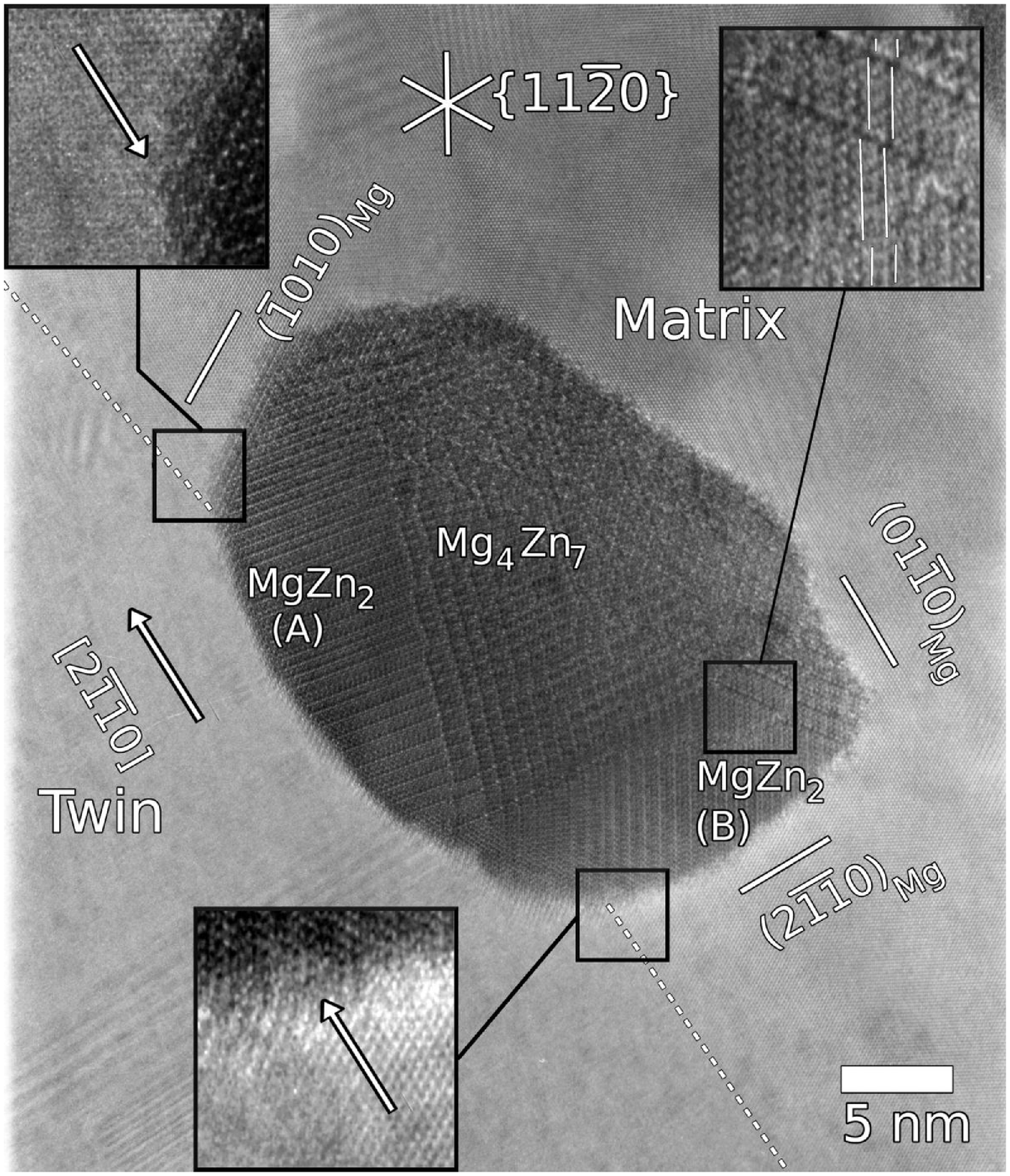}}

\hfill
\subfigure[Mg Matrix \label{fft-1}]{\fig[5.4cm]{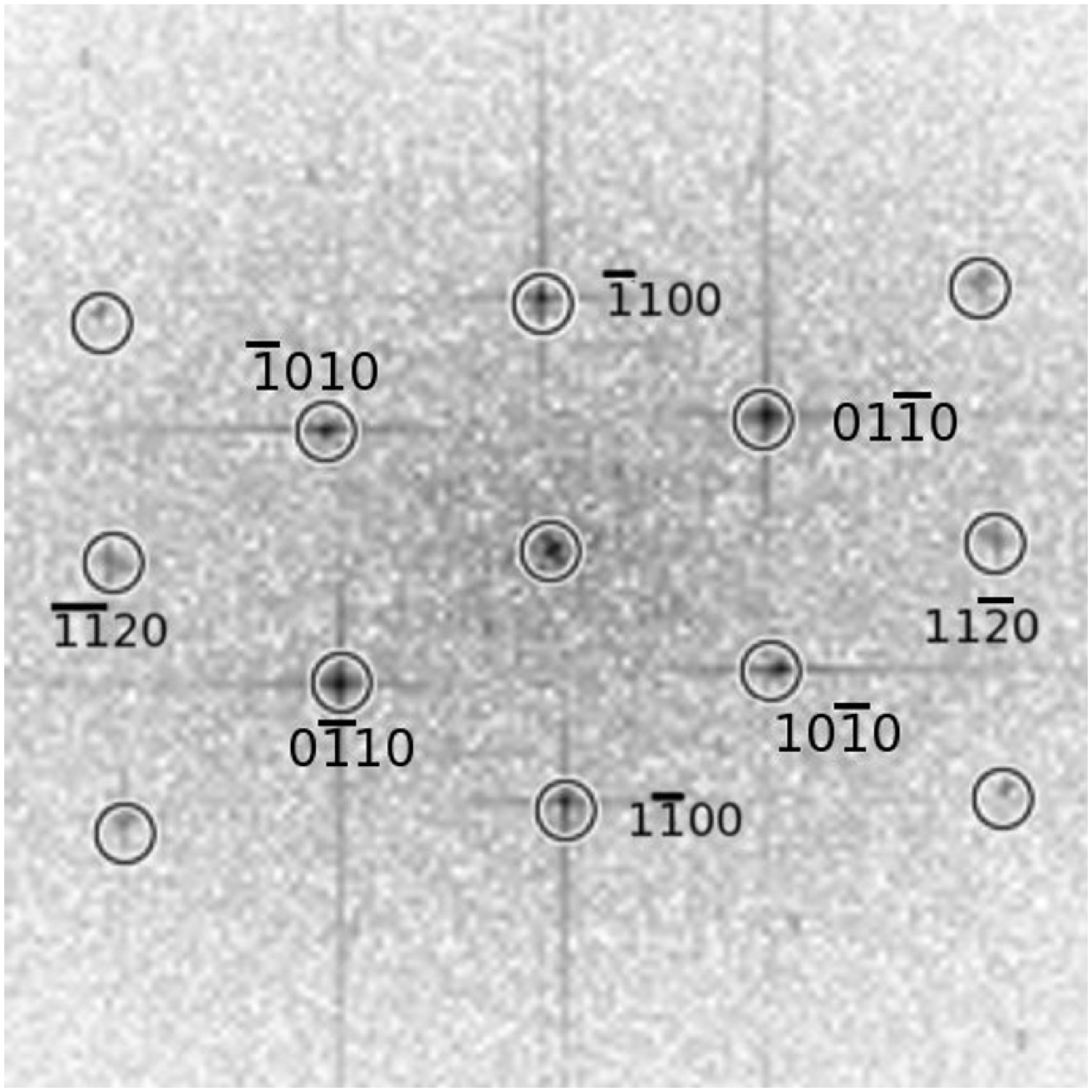}}\hfill
\subfigure[Central \ce{Mg4Zn7} region \label{fft-2}]{\fig[5.4cm]{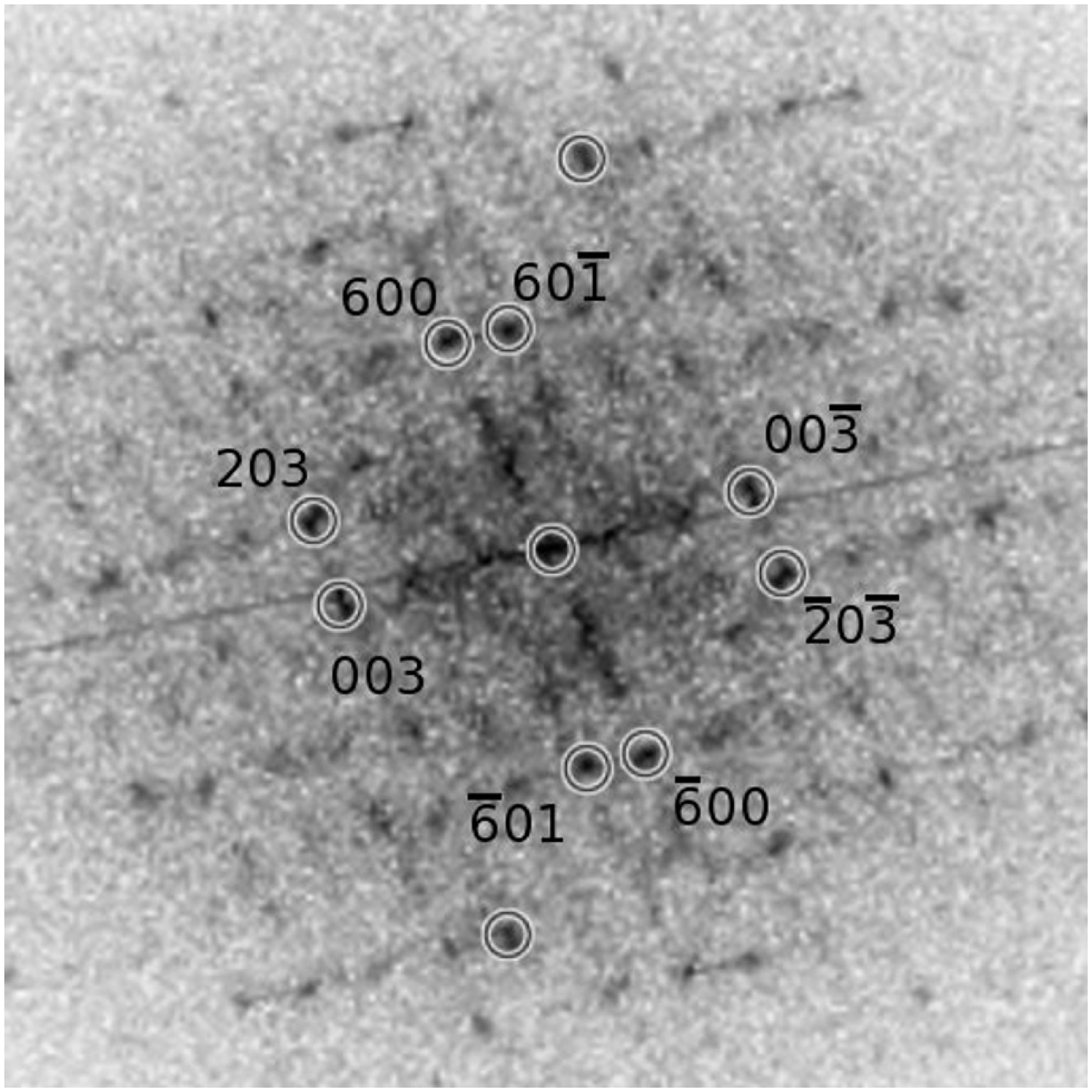}}\hfill\

\hfill \subfigure[hexagonal Laves phase adjacent to $A$ \label{fft-3}]{\fig[5.4cm]{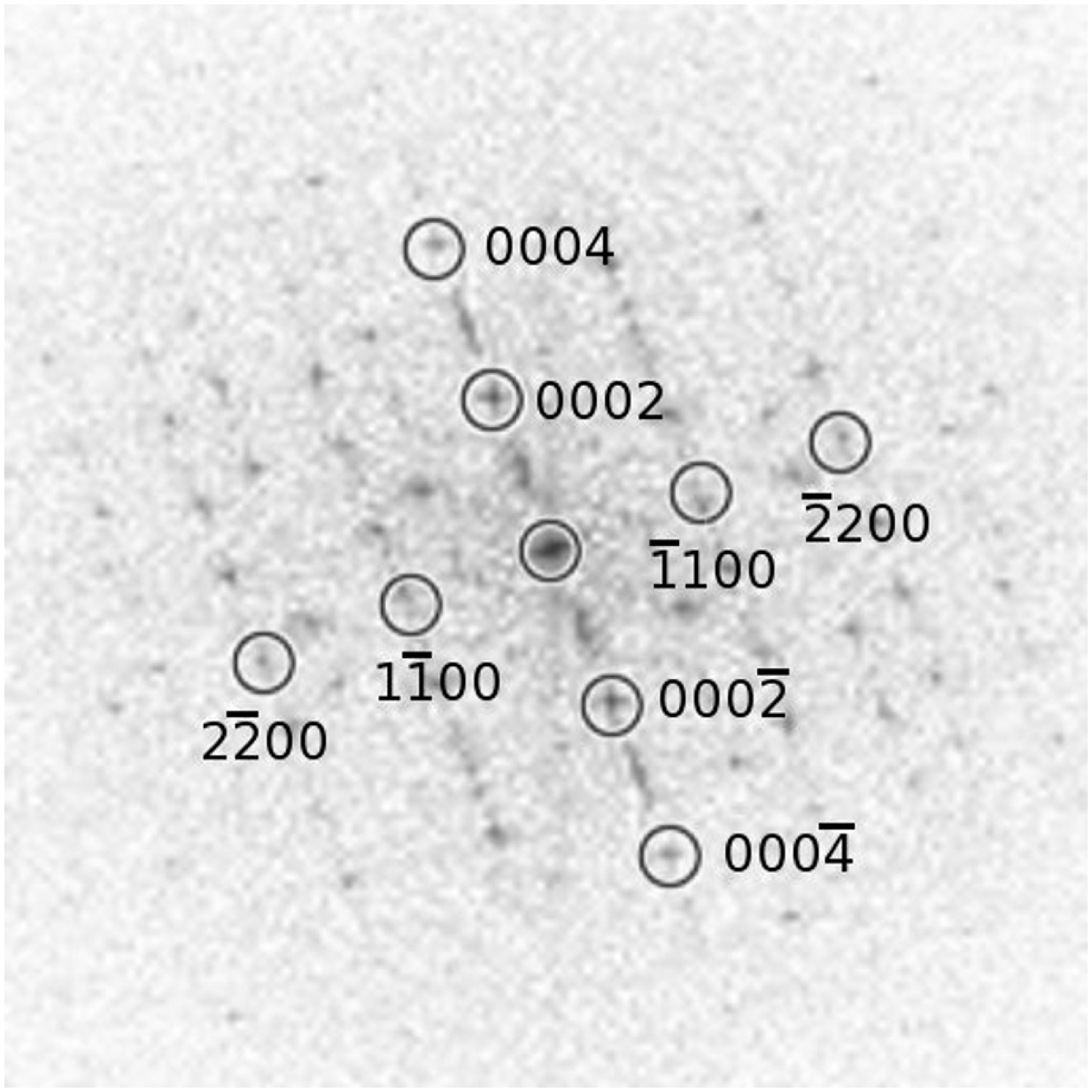}}\hfill
\subfigure[hexagonal Laves phase adjacent to $B$ \label{fft-4}]{\fig[5.4cm]{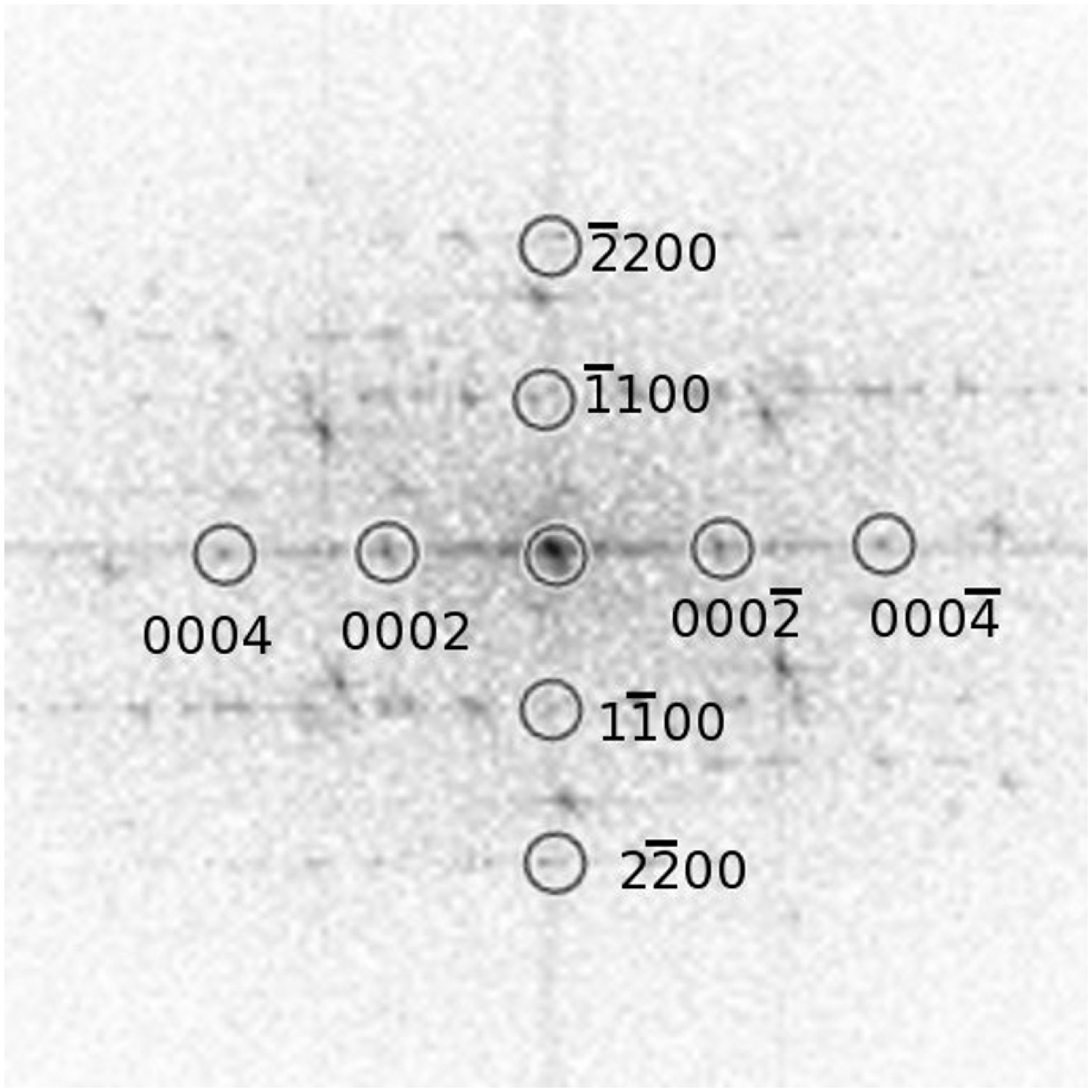}}\hfill\

\caption{High resolution TEM image of a twin boundary precipitate containing both \ce{Mg4Zn7} and \ce{MgZn2} phases. 
a) Micrograph showing  central and upper right regions belonging to the \ce{Mg4Zn7} phase.
The left \((A)\) and lower \((B)\) regions consist of the \ce{MgZn2} phase.
The enlarged insets show where the precipitate intersects the twin-boundary. 
Fourier transforms of the matrix, monoclinic and hexagonal Laves phases are given in (b)-(e). 
 \label{hrtem-1}}
\end{center}
\end{figure}

An example of such a dual-phase precipitate is presented in the HRTEM micrograph of Figure~\ref{hrtem-1}. 
The matrix is viewed along the hexagonal axis, as evidenced by FFT in Figure~\ref{fft-1}, while the twin region is close to the \( [01\overline{1}0] \) zone. 
The image was from a relatively thicker portion of the foil and was obtained close to the Scherzer defocus, which is about -45nm.
The position of the twin boundary is indicated by the dashed lines and can also be seen in the enlarged insets showing where the  precipitate is in contact with the twin boundary. 
The central region of the precipitate 
is in contact with the matrix and belongs to the \ce{Mg4Zn7} phase with the monoclinic
 structure clearly evident, and confirmed by indexing of the FFT shown in Figure~\ref{fft-2}.
This central region of the precipitate has the orientation relationship 
\( [0001]_{\mathrm{Mg}} \parallel [010]_{\ce{Mg4Zn7}} \); 
\( (10\overline{1}0)_{\mathrm{Mg}} \parallel  (201)_{\ce{Mg4Zn7}} \) with respect to the matrix. 
In all such precipitates, the monoclinic phase was found to form with the \([010]_\ce{Mg4Zn7}\) axis aligned with the [0001] direction  of either the matrix or the twin, with no apparent preference for either region. 

In the left \((A)\) and lower \((B)\) regions of this precipitate, the structure looks changed from the monoclinic in between. FFTs were obtained from these, shown in Figures~\ref{fft-3} and \ref{fft-4}. Both of these could be indexed to the \ce{MgZn2} phase in zone axis \( [11\overline{2}0] \). Thus regions A and B in Figure~\ref{hrtem-1} are two differently oriented variants of the \ce{MgZn2} phase, which are in contact with both the matrix and the twin. 

The FFT for the monoclinic phase (Figure~\ref{fft-2}) contains a number of streaks lying parallel to the (100) and (001) planes.
Similar features have been reported for rod-like \betap precipitates of the monoclinic phase and the presence of a significant population of stacking faults or other planar defects was inferred \cite{Gao2007,Singh2007}.

\begin{figure}
\begin{center}
\hfill
\subfigure[\label{fig-stereo1}]{\fig[0.48\textwidth]{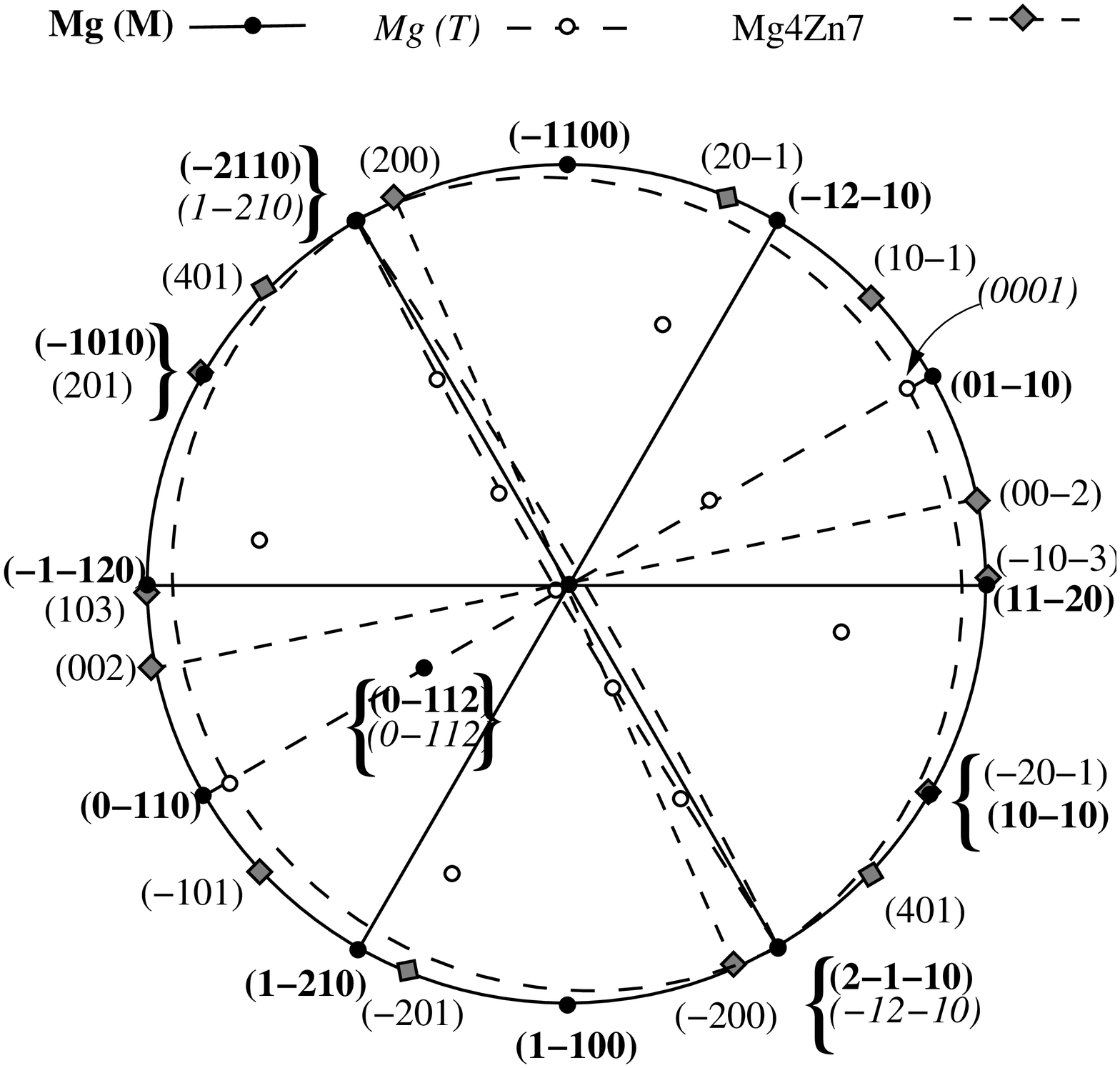}}
\hfill
\subfigure[\label{fig-stereo4}]{\fig[0.48\textwidth]{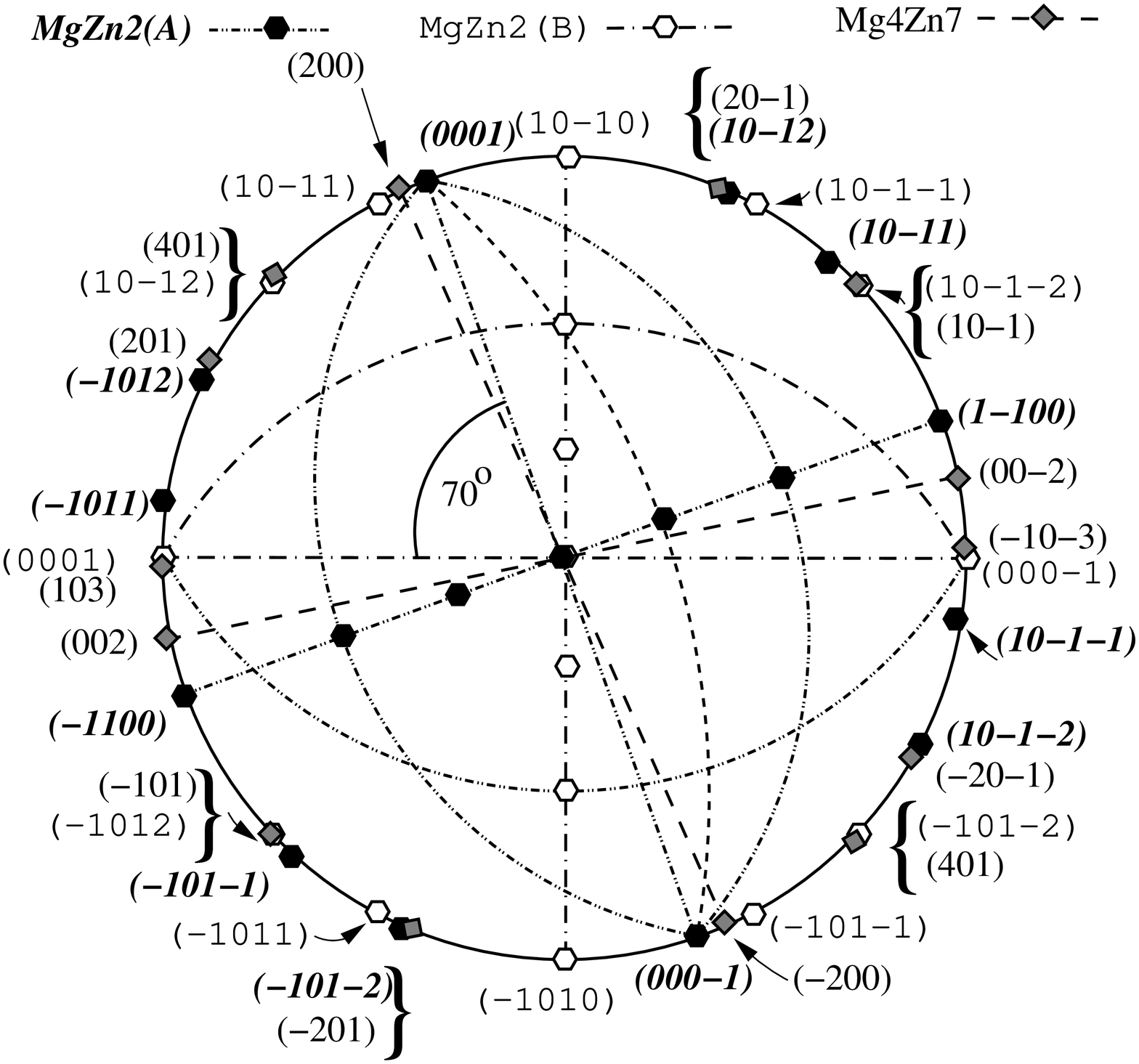}}
\hfill\

\hfill
\subfigure[\label{fig-stereo2}]{\fig[0.48\textwidth]{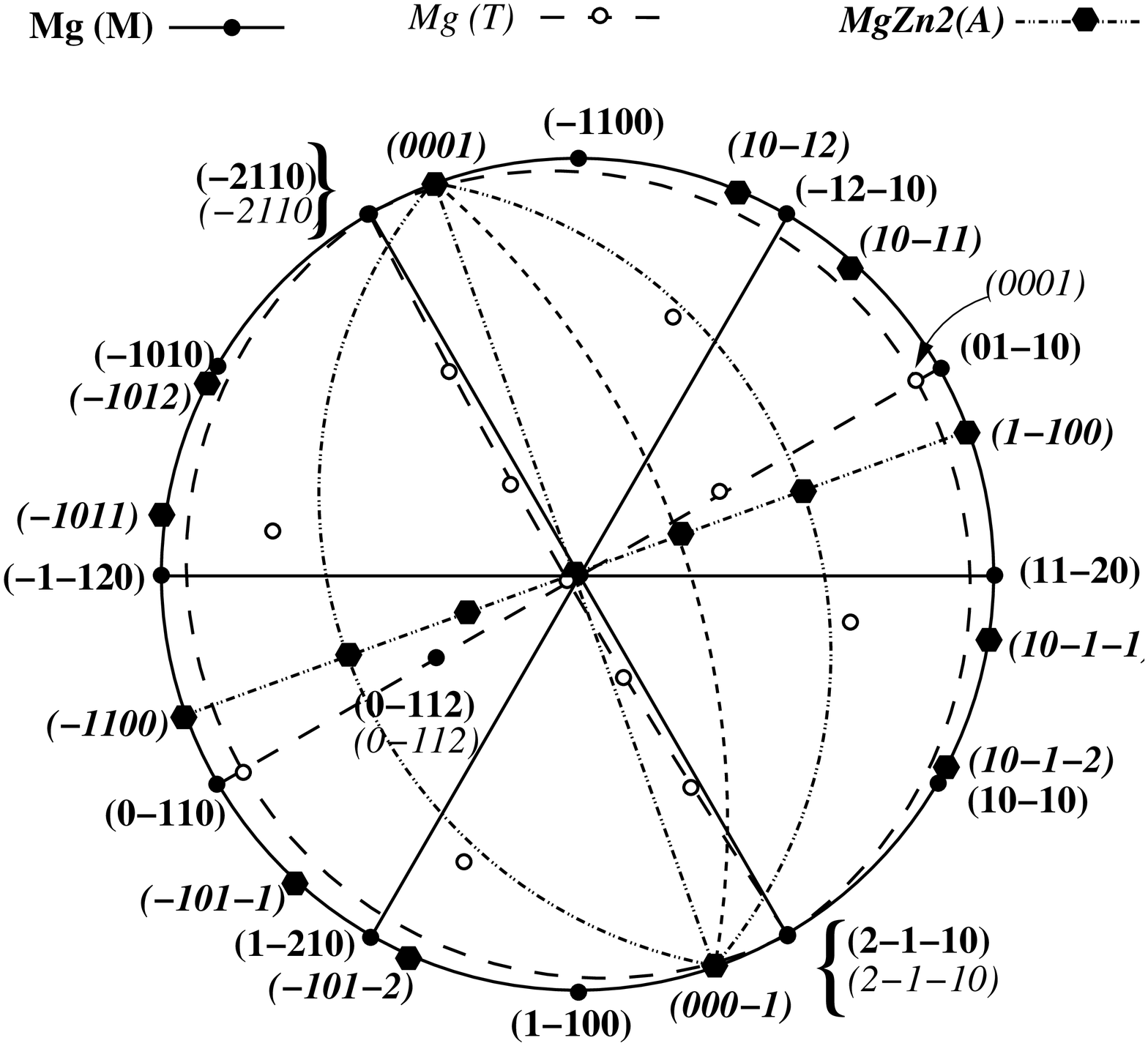}}
\hfill
\subfigure[\label{fig-stereo3}]{\fig[0.48\textwidth]{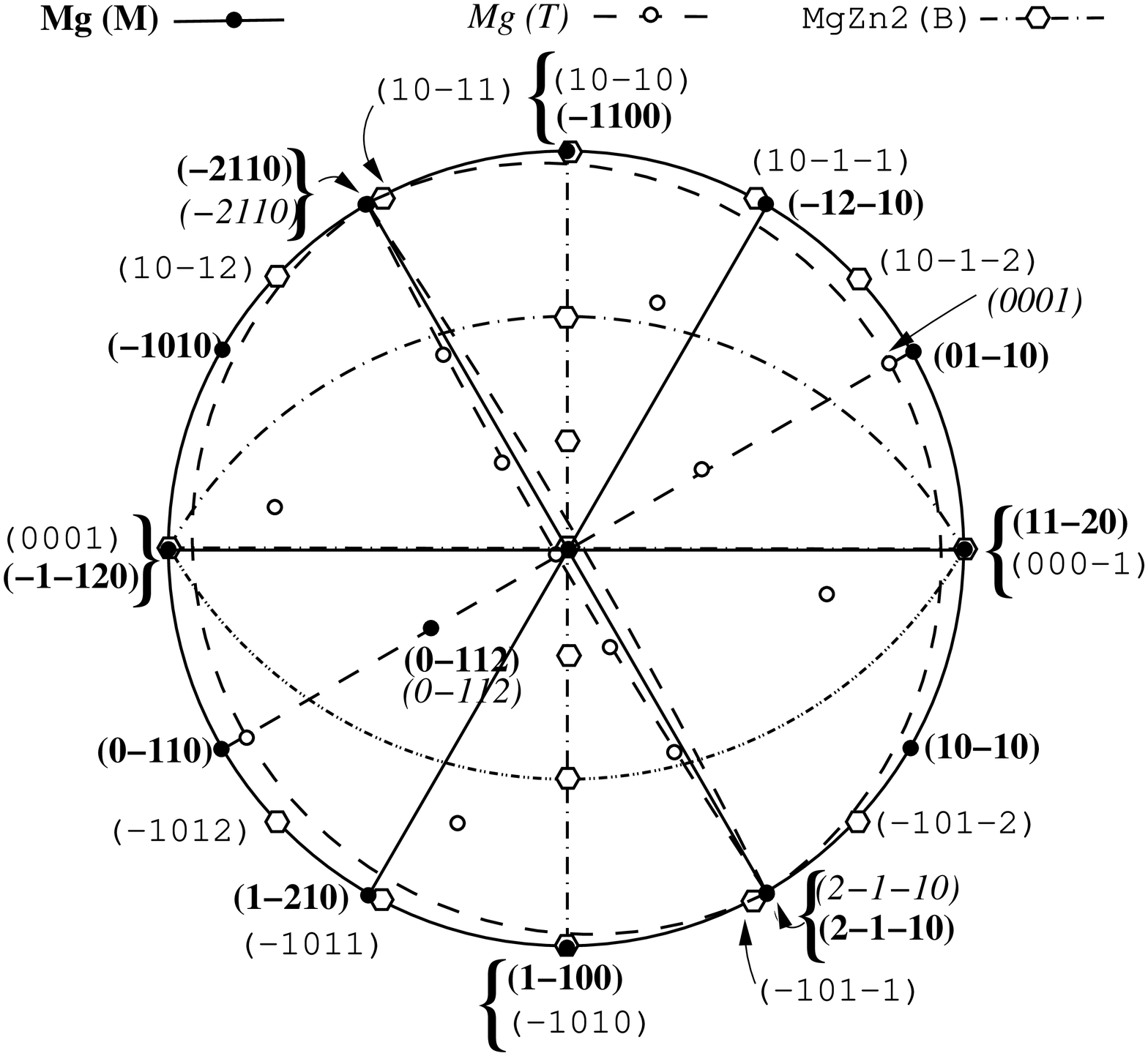}}
\hfill\

\caption{Stereographic projections illustrating the orientations of the phases in Figure \ref{hrtem-1}.
The centre of  each figure is aligned along the $[0001]_{M}$ $\parallel$ $[010]_\ce{Mg4Zn7}$ $\parallel$  
$[\overline{1}2\overline{1}0]_\ce{MgZn2}$.
Figure a) shows the matrix (M), twin (T) and monoclinic \ce{Mg4Zn7} phase. 
b) shows the stereograms for the two \ce{MgZn2} regions along with that of \ce{Mg4Zn7}.
c) shows the matrix and twin along with \ce{MgZn2} region A and 
d)  shows the matrix and twin and \ce{MgZn2} region B. 
\label{fig-stereo}}
\end{center}
\end{figure}

Based on these orientations, stereograms were plotted to see the ORs between the phases. The stereogram in Figure~\ref{fig-stereo1} shows the ORs between the matrix, the twin and the \ce{Mg4Zn7} phase.
As mentioned before, 
the (201) plane of the \ce{Mg4Zn7} phase is along the $(\overline{1}010)$ plane of the matrix ($M$).
Another close match is \((103)_\ce{Mg4Zn7} \parallel (\overline{11}20)_{M}\). 
No perfect match between \ce{Mg4Zn7} and the twin ($T$) was found. 
However, a close match is 
\((20\overline{4})_\ce{Mg4Zn7} \parallel (0001)_T\).

Figure~\ref{fig-stereo4} shows stereograms of \ce{Mg4Zn7} phase and two orientations, A and B, of \ce{MgZn2} phases which grow over it. 
A \( \{11\overline{2}0\}\) plane of the \ce{MgZn2} phase in each orientation is parallel to the (010) plane of the \ce{Mg4Zn7} phase. 
The hexagonal plane of \ce{MgZn2}(A) matches with $(200)_\ce{Mg4Zn7}$ within a few degrees. 
A near perfect match is 
\((10\overline{1}2)_\ce{MgZn2}(A)\parallel (20\overline{1})_\ce{Mg4Zn7} \). 
In the case of \ce{MgZn2} in orientation B, which shows a well-known orientation with the matrix,
its hexagonal axis is close to $(103)_\ce{Mg4Zn7}$.
It also has close matches between 
$ (10\overline{12})_{\ce{MgZn2}(B)} \parallel (10\overline{1})_\ce{Mg4Zn7}$
and
$ (10\overline{1}2)_{\ce{MgZn2}(B)} \parallel (401)_\ce{Mg4Zn7}$.

Both variants of the hexagonal Laves phase have the $[\overline{1}2\overline{1}0] _\ce{MgZn2}$ direction  
parallel to $ [0001]_\mathrm{Mg}$ with respect to the matrix.
This matching has been reported for intragranular \(\beta_1^\prime\) precipitates of the Laves phase in these alloys \cite{Takahashi1973,Mendis2009}. 
Figure~\ref{fig-stereo2} shows ORs of \ce{MgZn2} in orientation A with the matrix and the twin. 
\ce{MgZn2}(A) in fact shows no other significant relationship with the matrix or the twin. 
One  close match is 
\((\overline{1}010)_M\parallel(\overline{1}012)_{\ce{MgZn2}(A)}\). 

A definite OR occurs between \ce{MgZn2} in orientation B and the matrix and twin, shown in Figure~\ref{fig-stereo3}. 
As known, \( (\overline{1}\overline{1}20) \parallel (0001)_{\ce{MgZn2}(B)} \)
and \( (\overline{1}100)_M \parallel (\overline{1}010)_{\ce{MgZn2}(B)} \). 
No simple OR was found with the twin. 
The ORs between the phases are summarised in Table~\ref{tab-ors}. 

\begin{table}
\renewcommand{\arraystretch}{1.3}
\begin{center}
\caption{Orientations of the Mg,\ce{MgZn2} and \ce{Mg4Zn7} phases present in Figure~\ref{hrtem-2}.
$\sim$ indicates planes which are approximately parallel.
The \ce{Mg4Zn7} phase has the orientation relationship:
\([0001]_\mathrm{Mg}\parallel[010]_\mathrm{\ce{Mg4Zn7}}\);
\((\overline{1}010)_\mathrm{Mg}\parallel(201)_\mathrm{\ce{Mg4Zn7}}\).
The two \ce{MgZn2} regions not share a rational, low-index orientation relationship with the matrix.
None of the precipitate phases is rationally oriented with respect to the twin region of the matrix. 
\label{tab-ors}}
\begin{tabular}{rcccc}
\hline
& Twin & Monoclinic & Laves(A) & Laves(B) \\
&         -Mg      & -\ce{Mg4Zn7} & -\ce{MgZn2} & -\ce{MgZn2} \\ \cline{2-5}
Matrix
& $[\overline{2}110]\parallel [\overline{2}110]$ 
& $[0001]\parallel[010]$	
& $[0001]\parallel [\overline{1}2\overline{1}0]$
& $[0001]\parallel [\overline{1}2\overline{1}0]$
\\
(Mg-)
&   $(0\overline{1}12) \parallel (0\overline{1}12)$ 
&$(\overline{1}010)\parallel(201)$	
& $(\overline{1}010)\sim (\overline{1}012) $
&$(11\overline{2}0)\parallel(0001)$	
\\
Twin
& \dots
& $[01\overline{1}0]\sim [010]$
& $[01\overline{1}0]\sim [\overline{1}2\overline{1}0]$
& $[01\overline{1}0]\sim [\overline{1}2\overline{1}0]$
\\
(Mg-)
&  
& $(\overline{1}2\overline{1}2) \sim (\overline{4}01)$
& $(1\overline{2}11) \sim (0001) $
& $(\overline{2}110) \sim (10\overline{1}1)$
\\
Monoclinic
& \dots
& \dots 
& $ [010] \parallel [\overline{1}2\overline{1}0]$
& $ [010] \parallel [\overline{1}2\overline{1}0]$
\\
(\ce{Mg4Zn7}-)
&  
& 	
&$(20\overline{1})\sim  (10\overline{1}2)$
& $ (20\overline{2}) \sim 10\overline{1}\overline{2})$ 
\\
Laves(A)
& \dots
& \dots
& \dots
& $[\overline{1}2\overline{1}0] \parallel [\overline{1}2\overline{1}0]$
\\
(\ce{MgZn2}-)
&
&
&
& $(10\overline{1}2) \sim (10\overline{11})$
\\
\hline
\end{tabular}
\end{center}
\end{table}

\begin{figure}[htbp]
\begin{center}
\fig[11cm]{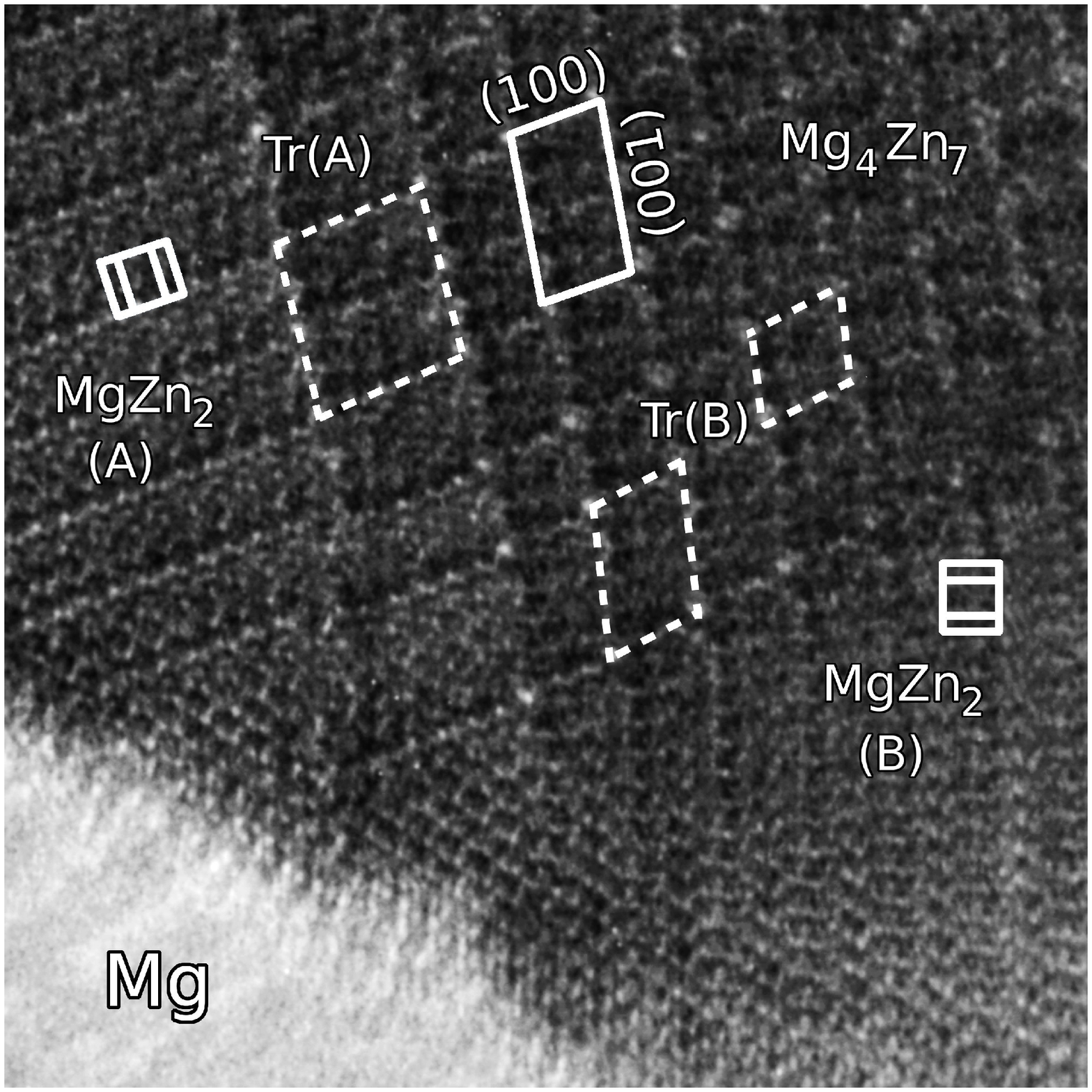}
\caption{An enlarged region of the twin-boundary particle shown in Figure~\ref{hrtem-1}. 
The central region (upper right in \ref{hrtem-1}(a) and FFT in \ref{fft-2})  has the \ce{Mg4Zn7} structure.
Two differently oriented variants of the \ce{MgZn2} phase are to the left and lower portions of the image ($A$ and $B$, respectively, with FFTs in \ref{fft-3},\ref{fft-4}) . 
The unit cells for the \ce{Mg4Zn7} and \ce{MgZn2} structures have been outlined on the image.  
Transition structures denoted \(Tr(A)\) and \(Tr(B)\) and indicated by dotted lines can be seen at the \ce{Mg4Zn7}-\ce{MgZn2} interfaces. 
\label{hrtem-2}
} 
\end{center}
\end{figure}

An enlarged micrograph of the same precipitate is shown in Figure~\ref{hrtem-2} to show the interfaces between the monoclinic and hexagonal Laves phases in greater detail.  
It is possible to identify complete unit cells  of the \ce{Mg4Zn7} structure (solid lines). 
The unit cells of the two differently oriented \ce{MgZn2} hexagonal Laves phase  structures 
(labeled \(A\)  and \(B\), respectively) are also indicated by solid lines. 
In between these three regions of central \ce{Mg4Zn7} structure and A and B, cells can be recognised which resemble those of the \ce{Mg4Zn7} structure but with different dimensions. 
Under the present imaging conditions the corners of the \ce{Mg4Zn7} unit cell, and the midpoints of the $(100)_\ce{Mg4Zn7}$ planes appear as brighter spots, corresponding to magnesium-rich sites within the structure. 
The spacing between these sites increases close to both interfaces with the Laves phase indicating a change in the lattice spacing in either the [100] or the [001] direction.
Aside from the increased spacing between the magnesium-rich sites, there did not appear to be any significant changes to the structure.
These regions are considered as transition structures in between \ce{Mg4Zn7}-\ce{MgZn2}, which are related to but distinct from the \ce{Mg4Zn7} structure. 
These are indicated by dashed lines, denoted \(Tr(A)\) and \(Tr(B)\), and described in greater detail. 

The presence of these transition structures suggests that the structure has undergone not a sudden but a gradual transition between the \ce{Mg4Zn7} and \ce{MgZn2} structures. 
Transition structures with monoclinic cells, (labelled $Tr(A)$ and $Tr(B)$) can be identified near the \ce{Mg4Zn7}-\ce{MgZn2} interfaces and are indicated by dashed  lines. 
The spacing between magnesium-rich sites at $Tr(A)$ was
2.6\,nm and 2.42\,nm in the $[100]_\ce{Mg4Zn7}$ and $[001]_\ce{Mg4Zn7}$ directions respectively.
Two cells are outlined in $Tr(B)$ where
the spacing remained $c$= 1.48\,nm parallel to $[001]_\ce{Mg4Zn7}$, and was measured at 1.59\,nm and 2.3\,nm  
parallel to $[100]_\ce{Mg4Zn7}$ in the two cells illustrated by the dashed lines.

\begin{figure}
\begin{center}
\hfill
\subfigure[\label{monosima}]{\fig[5cm]{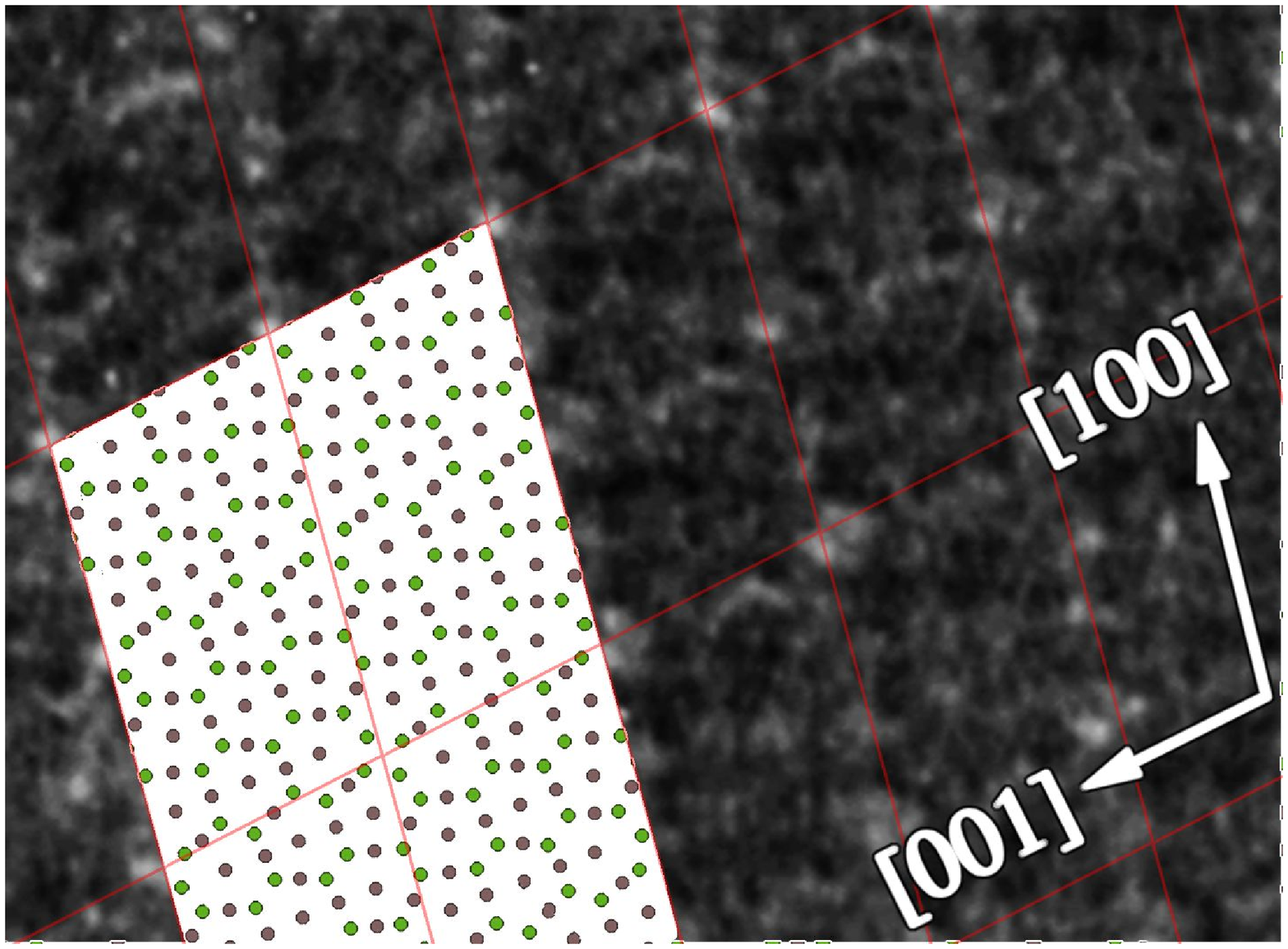}}
\hfill
\subfigure[\label{lavessima}]{\fig[5cm]{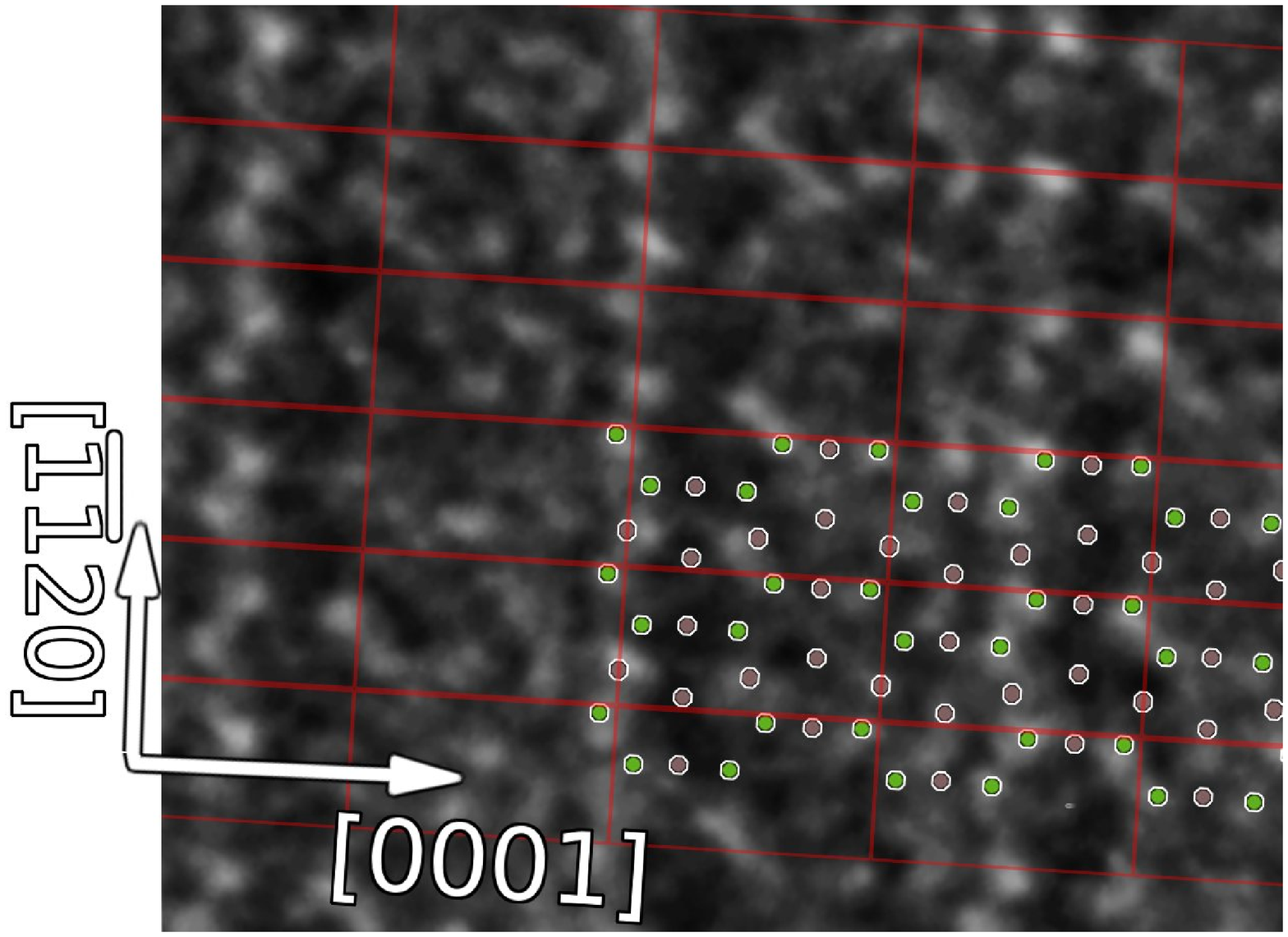}}
\hfill\

\hfill
\subfigure[\label{monosimb}]{\fig[5cm]{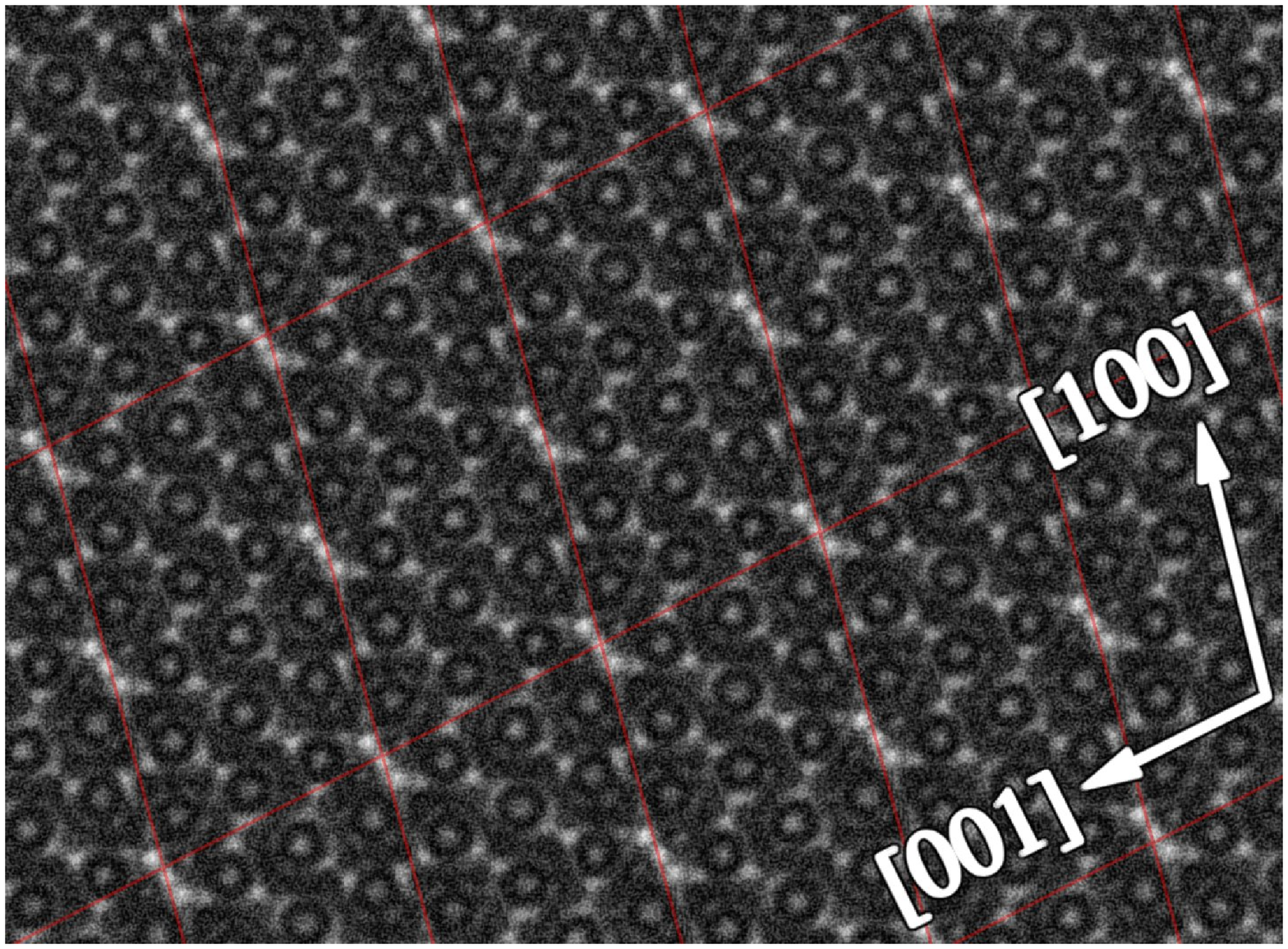}}
\hfill
\subfigure[\label{lavessima}]{\fig[5cm]{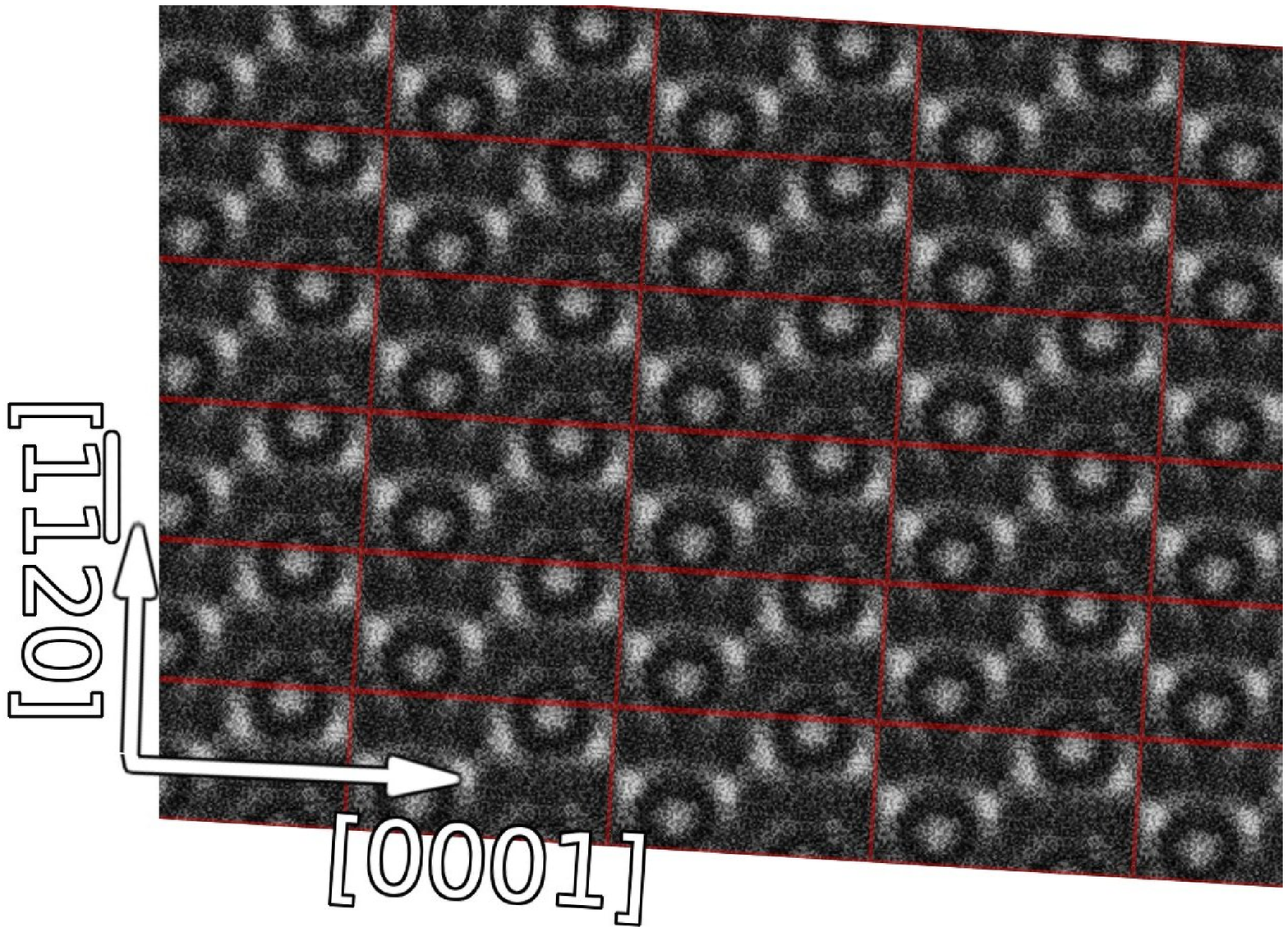}}
\hfill\

\caption{Multi-slice simulations for the (a,c) monoclinic \ce{Mg4Zn7} and (b,d) \ce{MgZn2} structures.
a and b) show the atomic sites and unit cells for the two phases overlaid over the experimental images. 
The images were taken from Figure~\ref{hrtem-2} close to the outlined unit cells for the monoclinic phase and region B of the Hexagonal phase.
c and d) are multi-slice simulations for a defocus of 75\,nm and a specimen thickness of 73\,nm. 
\label{mono-laves-sim}}
\end{center}
\end{figure}

The HRTEM images were compared with multi-slice simulations of the \ce{Mg4Zn7} and \ce{MgZn2} phases produced using JEMS software \cite{StadelmannJems1987} and are presented in Figure~\ref{mono-laves-sim}.  
Image simulations were produced for defocus values from  0 to 120\,nm and foil thicknesses up to 80\,nm. 
The region of the HRTEM image selected was that indicated in Figure~\ref{hrtem-2} by the outline of the relevant unit cell. 
Figures~\ref{mono-laves-sim}(a,b) show experimentally obtained images, overlaid with projected atomic positions and the unit cell boundaries within the monoclinic and the hexagonal Laves phases, respectively. 
Multi-slice simulation for these structures are shown in Figure~\ref{mono-laves-sim}(c,d). 
The simulation shown are for a defocus of 75\,nm, with a specimen thickness of 73.6\,nm.

In both experimental and simulated images of the monoclinic phase, the most prominent features are sites of strong contrast at the corners of the unit cell and at the $\frac{1}{2}a$ sites of the monoclinic unit cell.
These are spaced approximately 12.5\,nm in the [100] direction, and 14.5\,nm in the [001] direction, 
as expected for the \ce{Mg4Zn7} structure.

These strongly contrasting sites present in the \ce{Mg4Zn7} phase were not present for the hexagonal phase. 
however, in other respects the simulations are similar, owing to the presence of similarly co-ordinated atoms in each structure.
The Laves phase simulation shows a good qualitative match with the experimental image in terms of atomic positions. 
Both experimental and simulated images show a zig-zagged pattern of stronger contrast centred on the basal planes for $z=0$.
In the experimental image, this feature was strongly enhanced for for every 1.5 unit cell heights.
This feature was thought to be due to ordering on this plane, which is dealt with in a companion work \cite{Singh2009coexist}.

\section{Discussion}

The present work has shown that both \ce{Mg4Zn7} and \ce{MgZn2} phases form along deformation twin-boundaries in Mg-Zn-Y and that the \ce{Mg4Zn7} and \ce{MgZn2} phases can co-exist as distinct regions within a single precipitate. 
In agreement with earlier studies \cite{Singh2007, WeiPrec1995}, the precipitates showed no evidence of definite yttrium enrichment and are therefore taken to consist of  binary Mg-Zn phases. 
Microdomain structures have been reported in \betap precipitates \cite{Gao2007,Singh2007} and
co-existence of the \ce{Mg4Zn7} and \ce{MgZn2} phases has recently been indicated in \betap rods \cite{Singh2009coexist}.
However, in the larger-scale twin-boundary precipitates examined herein it has been possible to identify 
distinct regions of nanometer scale belonging to each phase and to examine the \ce{Mg4Zn7}-\ce{MgZn2} interfaces.

Two features deserve particular note; the orientation relationship between the two precipitate phases and
the presence of monoclinic transition structures at the interface between \ce{Mg4Zn7} and \ce{MgZn2}. 
The significance of both features will be discussed in terms of the structural similarities between the \ce{Mg4Zn7} and \ce{MgZn2} phases.

\subsection{Structural and orientation relationships between \ce{Mg_4Zn_7} and \ce{MgZn_2}  phases \label{mono-laves-str}}

\begin{figure}
\begin{center}
\hfill
\subfigure[\label{fig-rhomb-mono}]{\fig[5cm]{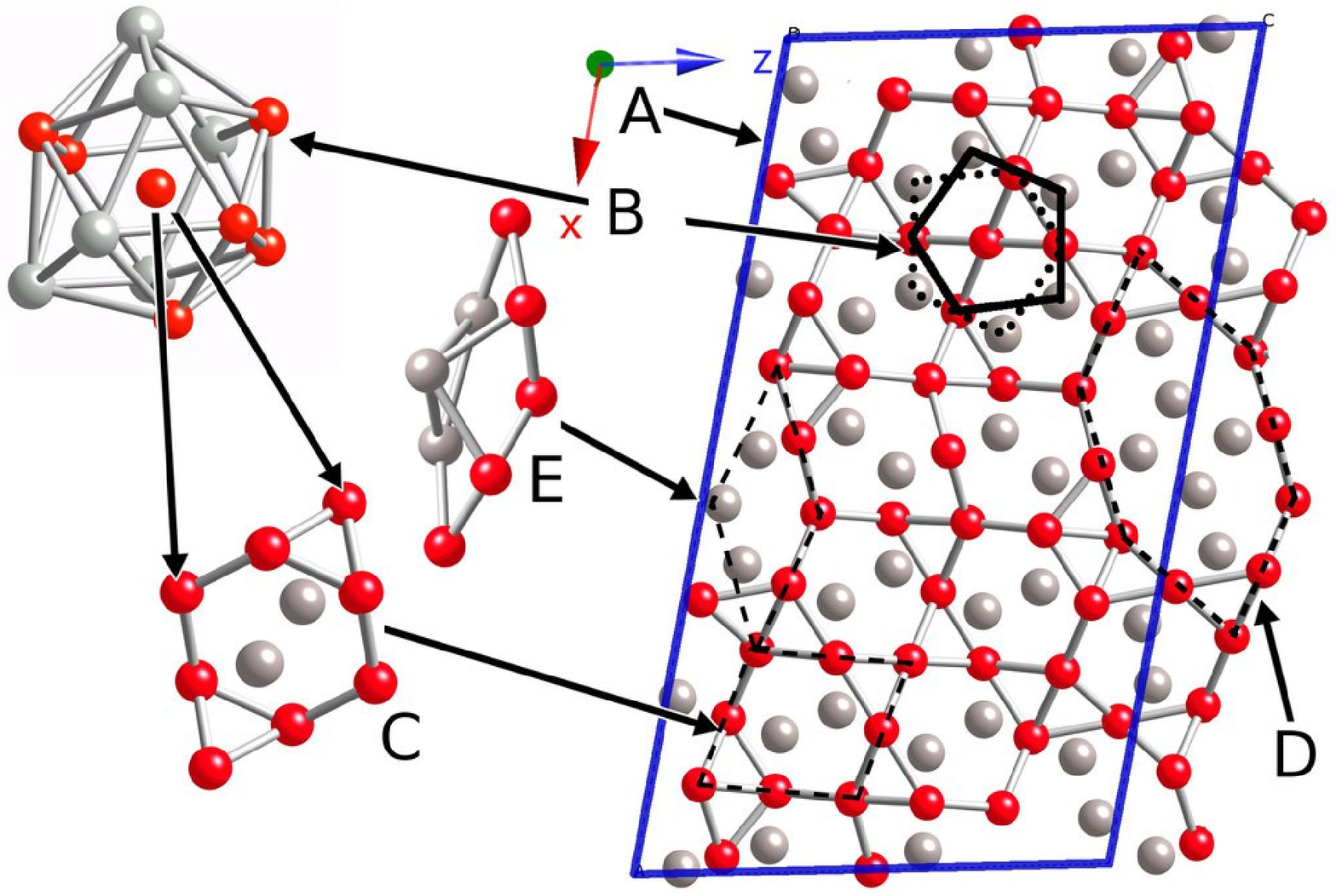}}\hfill
\subfigure[\label{fig-rhomb-laves}]{\fig[6cm]{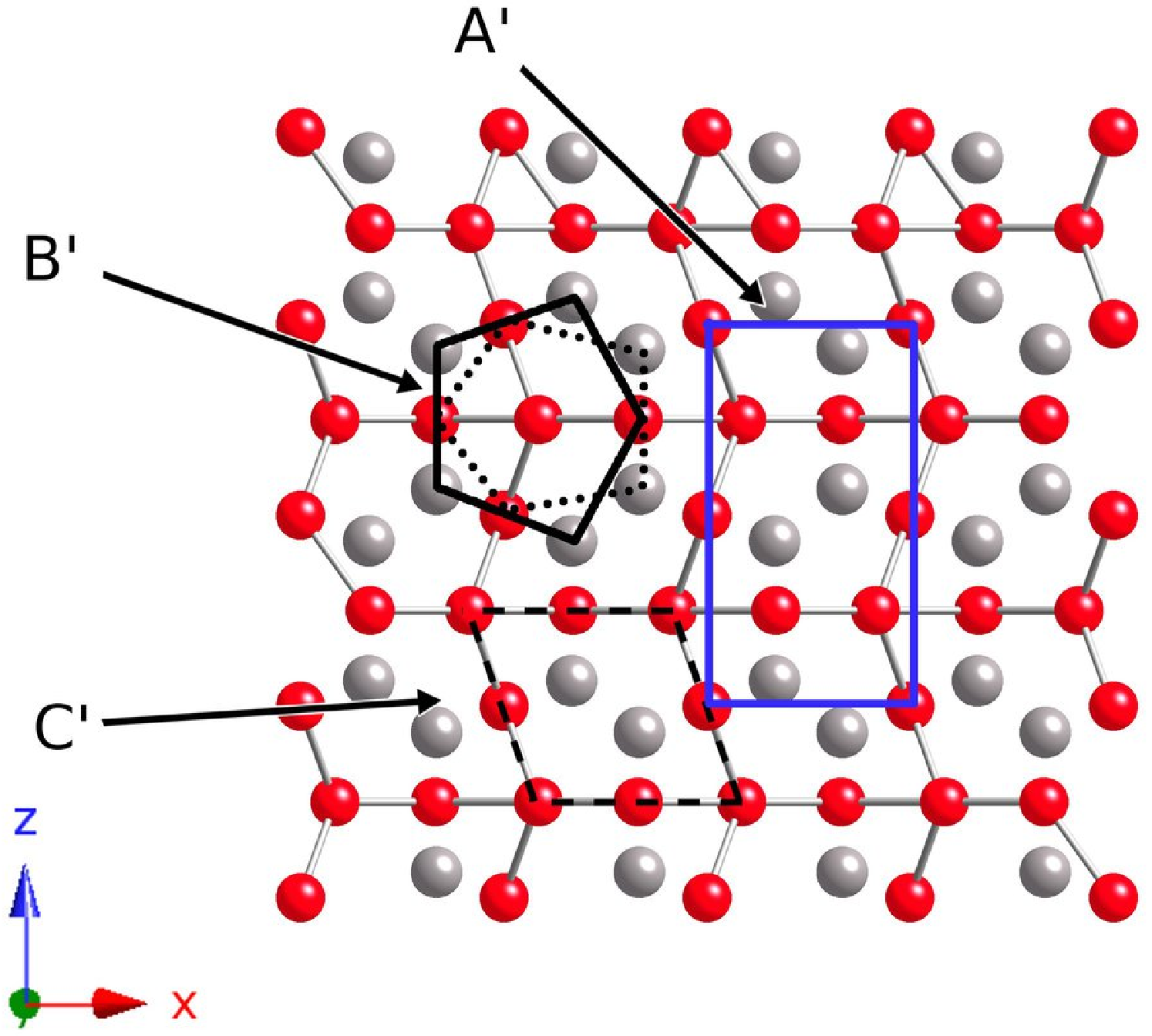}}
\hfill\

\subfigure[]{\fig[4.8cm]{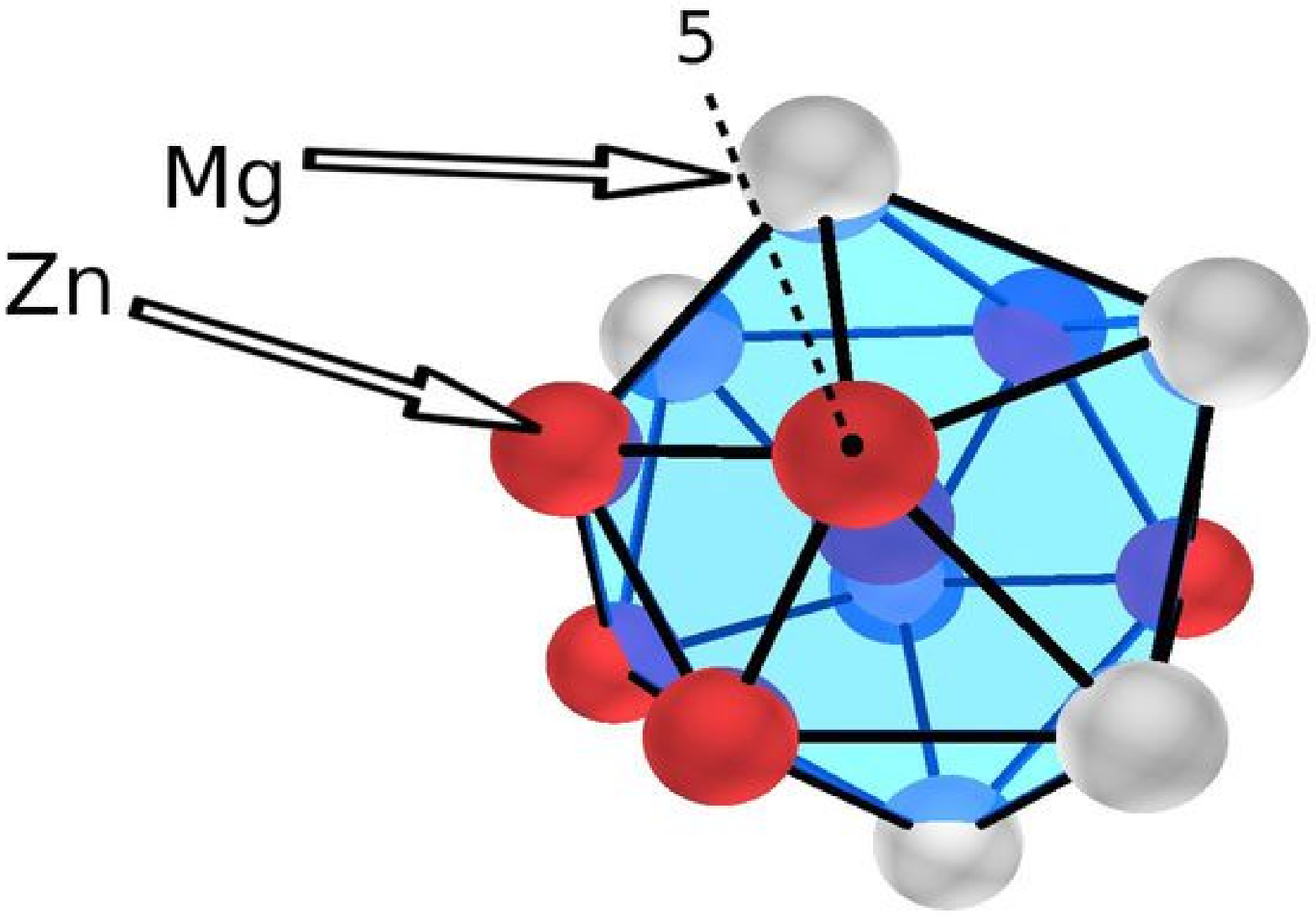}}
\caption{\label{fig-rhomb}
Structures of the (a) monoclinic and (b) hexagonal Laves phases.
The unit cell boundaries are indicated by the thick, solid lines (A, A').
The structures contain icosahedral clusters, centred on zinc atoms as shown at B, B' with the two pentagonal arrangements illustrated. 
Points C and C' indicate the positions of ``thick'' rhombic unit (with acute angle $\sim72^o$), the corner atoms of which have icosahedral co-ordination. 
The monoclinic structure also contains hexagonal units (D) at the unit cell corners and $\frac{1}{2}a$ sites. 
These can be decomposed into ``thin'' rhombic units (E) ( acute angle $\sim36^o$), both of which are absent from the hexagonal phase.
One icosahedral unit is shown in (c) where it is slightly inclined away from a 5-fold axis (indicated).}
\end{center}
\end{figure}

The structures of the monoclinic and hexagonal Laves phases are illustrated in Figure~\ref{fig-rhomb}. 
Figure~\ref{fig-rhomb-mono} shows the monoclinic structure along a [010] axis,
while Figure~\ref{fig-rhomb-laves} shows the hexagonal Laves phase structure along the $[11\overline{2}0]$ axis. 
Unit cell boundaries for the two phases (A and A', respectively) are indicated by  thick lines. 
Both phases contain icosahedral clusters centred on zinc atoms (indicated at points B and B'. respectively), with pentagonal layers above and below these zinc atoms.
The upper and lower pentagonal layers are rotated 180\degree with respect to each other \cite{Yang1987}.

The structure can also be viewed as a series of face-sharing ``thick'' rhombic tiles,
with acute angle $\sim72^o$, with the vertices of the rhombic tiles occupied by  zinc atoms which form the vertices of the icosahedral coordinations. 
Thick rhombic tiles are outlined at points C and C', respectively. 
Each ``thick''  rhombic unit has a composition of \(\mathrm{Mg_2Zn_4}\) and a continuous patterning of such units satisfies the overall \ce{MgZn2} composition of the hexagonal Laves phase.
In the hexagonal Laves structure, the rhomboids are arranged with the acute angle alternating right to left.
In contrast, the monoclinic structure in has acute angles directed Right-Right-Left.

The monoclinic structure also contains hexagonal units at the corners and $a=\frac{1}{2}$ sites.
These can be further decomposed into ``thin'' rhombic tiles (D) (acute angle $\sim36^o$) which are absent from the hexagonal phase. 
Each such unit is zinc-deficient relative to the Laves phase composition and these serve to establish the \ce{Mg4Zn7} composition.

When formed in the matrix, the monoclinic and hexagonal phases adopt axial orientations in which the hexagonal axis of magnesium is parallel to \([010]_\ce{Mg4Zn7}\) and to one of the \(\langle11\overline{2}0\rangle_\ce{MgZn2}\) directions, respectively.
This permits good matching between the basal plane of Mg (0.520\,nm) with \((010)_\ce{Mg4Zn7}\) (0.524\,nm)  \cite{Yarmolyuk1975,Gao2007} and  \( \{11\overline{2}0\}_\ce{MgZn2}\) (0.261\,nm).
The change in crystallographic orientation across the twin boundary makes it impossible for an entire precipitate consisting of either \ce{Mg4Zn7} or \ce{MgZn2} to sustain such a relationship with both matrix and twin. 
The monoclinic phase for example (space group \#12, with \(\beta =102.5\degree\)) is not able to simultaneously adopt any favourable planar matching with both matrix and twin, as can be seen in the stereogram in Figure~\ref{fig-stereo1}.

The hexagonal Laves phase, however, is able to assume more favourable orientations across the twin boundary.
For precipitates of the hexagonal phase on the twin boundary, 
as in Figure~\ref{laves-tbp}, 
the phase is close to a previously-reported orientation to the matrix \cite{SinghTMS2008}, 
with \( (11\overline{2}0)_\mathrm{Mg} \parallel (0001)_\ce{MgZn2}\);
\((0001)_\mathrm{Mg}  \parallel (11\overline{2}0)_\ce{MgZn2}\).
In this orientation the  \( (0001)_\ce{MgZn2} \) planes  will also be near-parallel to the \((11\overline{2}0)\)   planes  of the twin. 

In the mixed phase precipitates, the monoclinic phase was observed with the previously reported orientation relationships to Mg for the \betap (\ce{Mg4Zn7}) rods formed in the matrix \cite{Gao2007,Singh2007} and could be oriented with respect to either the matrix or twin. 
For the example shown in Figures~\ref{hrtem-1} and \ref{hrtem-2} the orientation relationship was
\( [0001]_{\mathrm{Mg}}\parallel[010]_{\ce{Mg4Zn7}} \);
\( (10\overline{1}0)_{\mathrm{Mg}}\parallel(201)_{\ce{Mg4Zn7}} \) with respect to the matrix.

When the hexagonal Laves phase co-existed with the monoclinic structure, 
one region of the Laves phase (B) adopted the known orientation 
\( [0001]_\mathrm{Mg}\parallel [1\overline{2}10]_{\ce{MgZn2}} \);
\( (\overline{11}20)_\mathrm{Mg}\parallel (0001)_{\ce{MgZn2}} \) \cite{Gao2007,WeiPrec1995}.
The other hexagonal region (A) did not have closely matching lattice planes with respect to either matrix or twin and at first glance its presence appears difficult to explain.

\begin{figure}[htbp]
\begin{center}
\hfill
\subfigure[\label{CellMatchA}]{\fig[6cm]{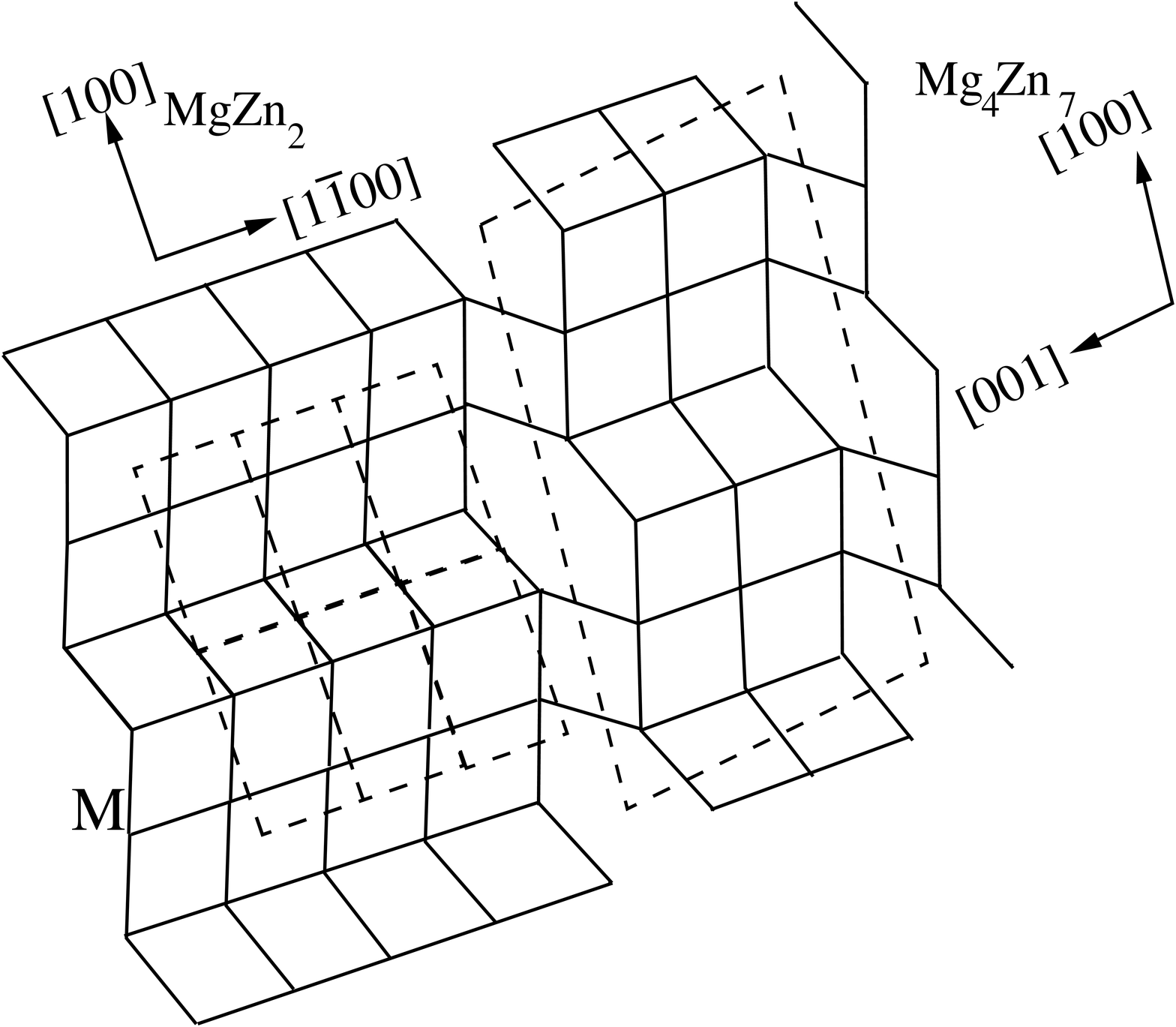}}\hfill
\subfigure[\label{CellMatchA}]{\fig[4.5cm]{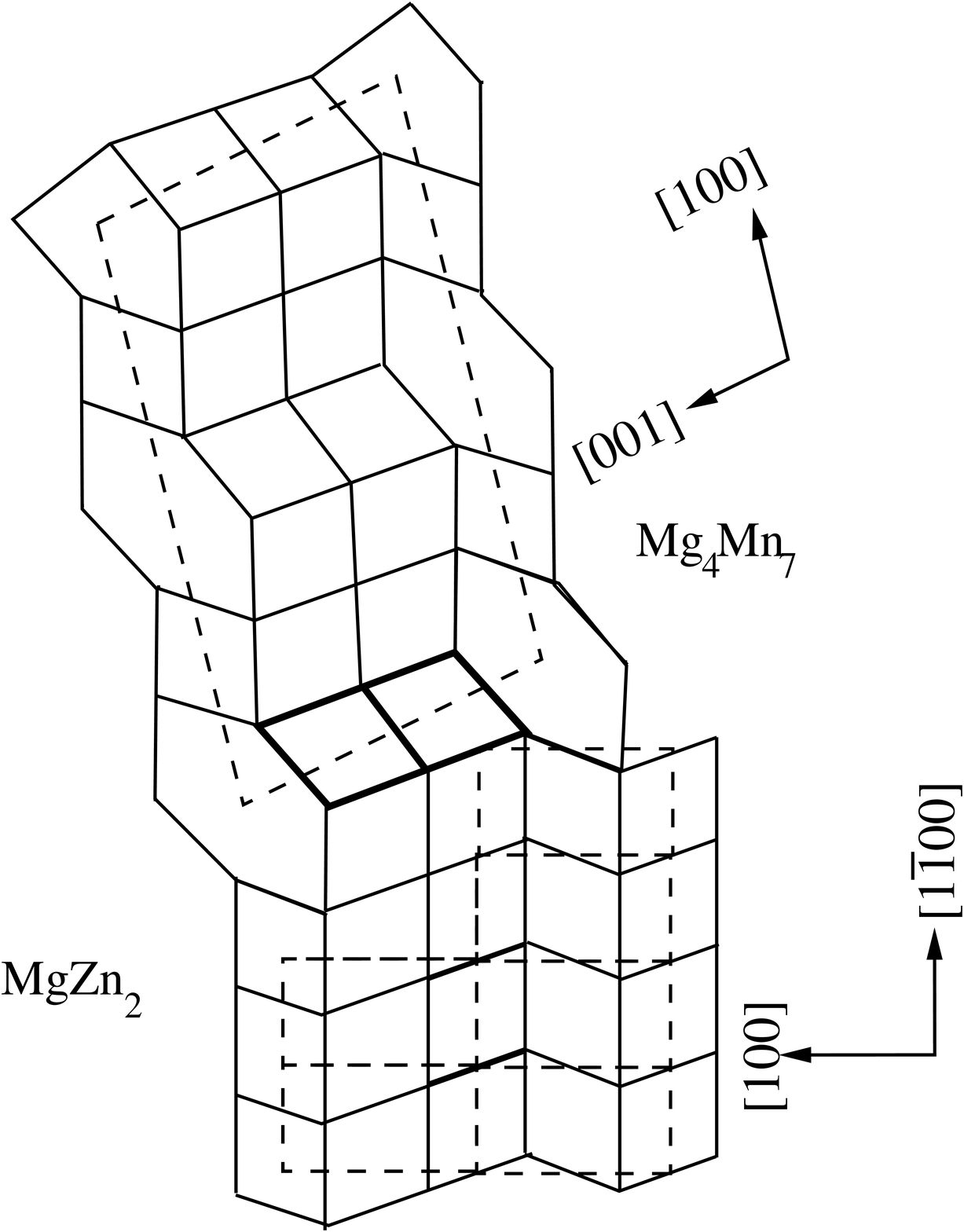}}
\hfill\

\caption{
Interface matching between the rhombic units of the \ce{Mg4Zn7} and \ce{MgZn2} phases for the alignment shown in Figure~\ref{hrtem-2}.
Rhombic units are shown with solid lines while the respective unit cells are indicated by dashed lines. 
a shows the matching between the \ce{Mg4Zn7} and hexagonal Laves phase as at $A$ in Figure~\ref{hrtem-2} and b) shows the corresponding match between the structures at $B$ in the same figure.
Note that although there is no clear relationship between the unit cell alignment of the phases, the rhombic unit remain closely aligned. \label{CellMatch}}
\end{center}
\end{figure}

It is possible, however, to demonstrate that the rhombic subunits present in \ce{Mg4Zn7} and \ce{MgZn2} remained aligned when \ce{Mg4Zn7} was in contact with either variant of \ce{MgZn2}. 
Figure~\ref{CellMatch}  presents schematics of the \ce{Mg4Zn7} and \ce{MgZn2} phases oriented with respect to one another as in Figure~\ref{hrtem-2} and shows that the orientation is not related by the unit cells, but through the constituent rhombic units. 
The unit cells for each structure are shown in solid lines, with the rhombic units outlined. 
It can be seen that the rhombic units within the \ce{Mg4Zn7} phase are accurately aligned with those of the two hexagonal Laves phase regions, and in principle at least, it appears possible to construct a well-matched interface for the two phases in this manner. 
It should be noted that in order to do so, a series of defects 
must be introduced parallel to the basal planes of the hexagonal Laves phase, with a spacing of $\sim$1.25\,nm.
Enhanced contrast with this spacing was observed in the experimental images for the hexagonal phase, which was not produced by the multi-slice simulations for the \ce{MgZn2} structure (Figure~\ref{mono-laves-sim}(b,d)). 
This would also be a source of stacking defects on {\it c}-planes observed in Figure~\ref{laves-tbp}.

The existence of such a relationship between the rhombic units and the absence of a unit-cell based correspondence with either monoclinic phase or magnesium  suggests that it is the rhombic units which are the determining factor in deciding the  orientation of the hexagonal Laves phase. 
The real interface is likely to be more complex than is set out in Figure~\ref{CellMatch}, 
particularly in terms of the introduction of defects and the presence of the transition structures. 
However, it appears likely that there will be good, possibly near-continuous, matching between the rhombic units making up the monoclinic and hexagonal phases.

No regularity was found in the stacking defects observed along the hexagonal axis of \ce{MgZn2} phase observed in Figure~\ref{laves-tbp}. 
Another type of planar defect parallel to \((\overline{1}101)_\ce{MgZn2}\) were pointed out in Figure~\ref{laves-tbp}. 
These defects are along only one direction and not on equivalent planes. These defects have a finite width, and across them, the (0001)$_\ce{MgZn2}$ planes were shifted by about half the interplanar distance. A rhombic unit can be defined in these defects. Therefore it can be assumed that these defects are also from the relationship of the \ce{MgZn2} phase with the \ce{Mg4Zn7}. 
In the stereogram of Figure~\ref{fig-stereo4}, a \{10$\overline{1}$1\}$_\ce{MgZn2}$ plane is nearly parallel to (200)$_\ce{Mg4Zn7}$ plane. In fact, when looked closely enough, such a defect is also observed in \ce{MgZn2} (B) in Figure~\label{hrtem1a}, in its upper portion. It is quite possible that the apparently single \ce{MgZn2} precipitates on the twin boundaries are also part of dual phase precipitates.

\subsection{Transition structures}

The existence of the transition structures at the \ce{Mg4Zn7}--\ce{MgZn2} interfaces (Figure~\ref{hrtem-2}) is a key observation.
These are structurally related to both phases, being constructed in a modular fashion from identical rhombic units  present in both \ce{Mg4Zn7} and \ce{MgZn2} phases, plus hexagonal units found in the \ce{Mg4Zn7} phase.
The crystal structure remains monoclinic but with enlarged unit cells, given in Table~\ref{cells2}. 

Such structures can be constructed from the standard \ce{Mg4Zn7} structure by extending the tiling of rhombic units. 
If the unit cell is extended by the addition of two rows of rhombic tiles in the  [001] directions it corresponds to transition structure \(Tr(A)\) in Figure~\ref{hrtem-2} with the 
 \ce{Mg4Zn7}-\ce{MgZn2} interface parallel to the \((001)_\ce{Mg4Zn7}\) plane. 
The cell can also be extended in the [100] direction,  
by the addition of two rhombic subunits,  
 corresponding to one of the transition structures, $Tr(B)$, at the \ce{Mg4Zn7}-\ce{MgZn2} interface parallel to the  \((001)_\ce{Mg4Zn7}\) plane. 

\begin{figure}
\begin{center}

\hfill
\subfigure[\label{extended001a}]{\fig[1.57in]{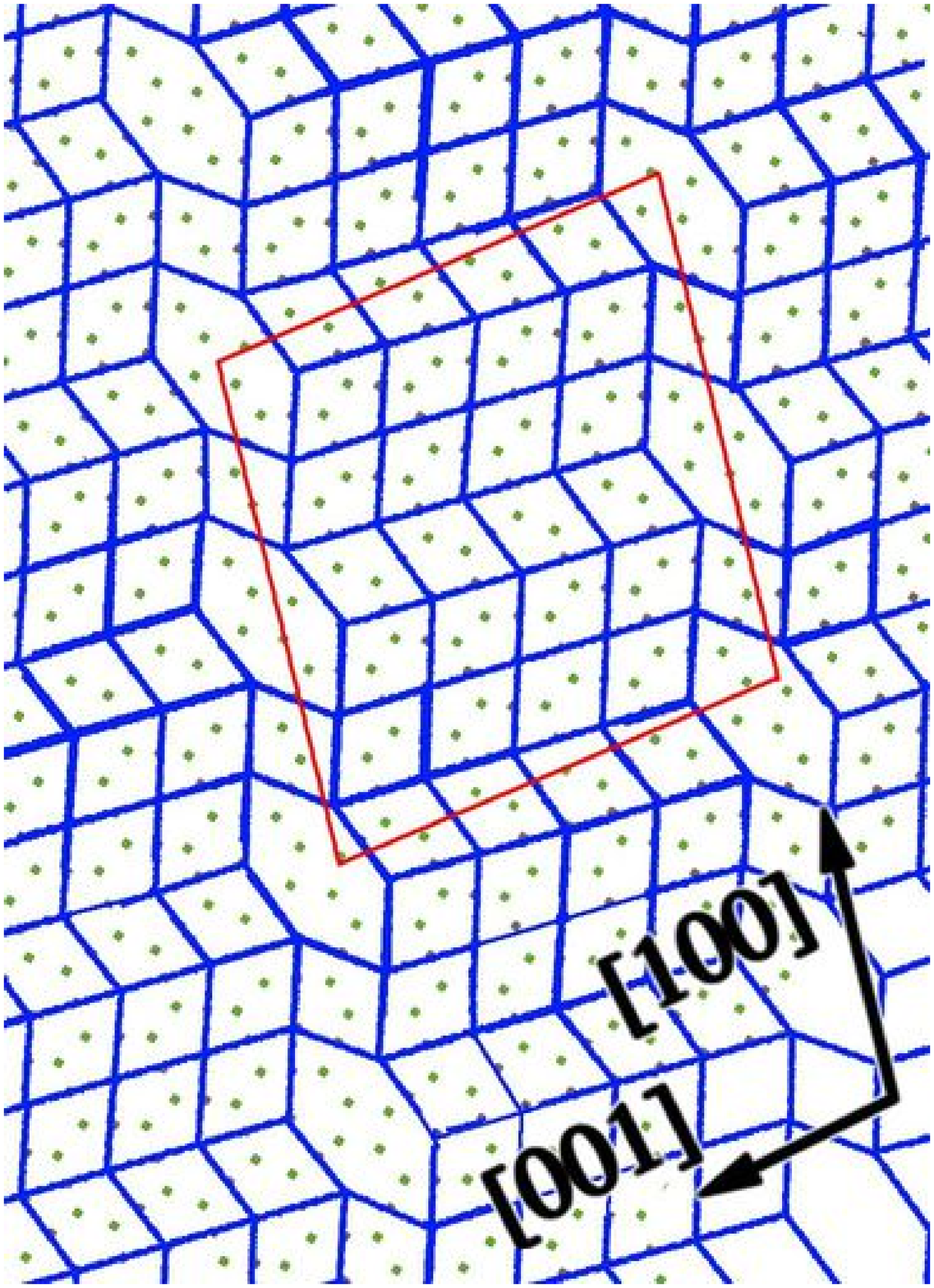}}
\hfill
\subfigure[\label{extended100a}]{\fig[1.8in]{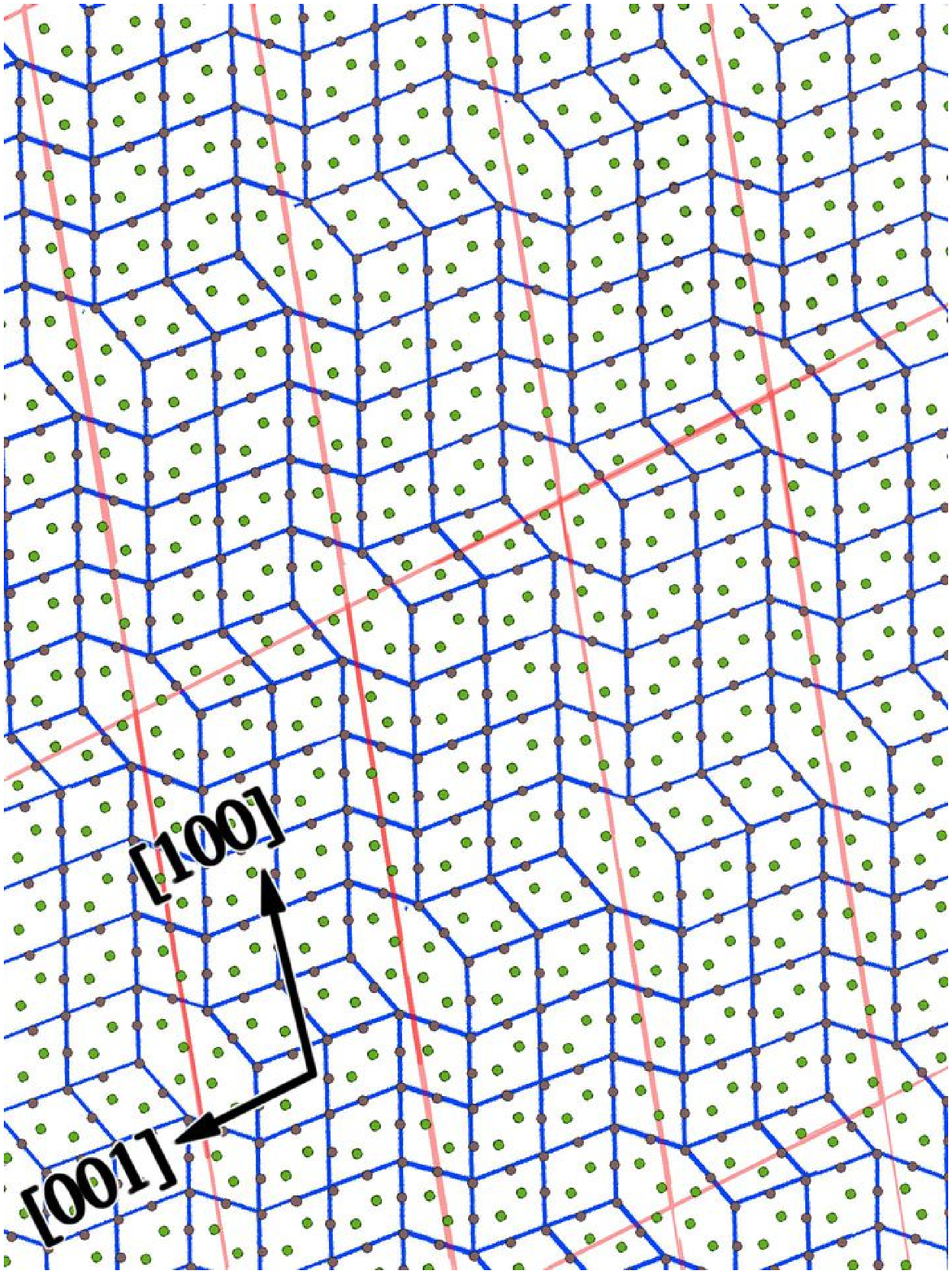}}
\hfill\

\hfill
\subfigure[\label{extended001b}]{\fig[1.57in]{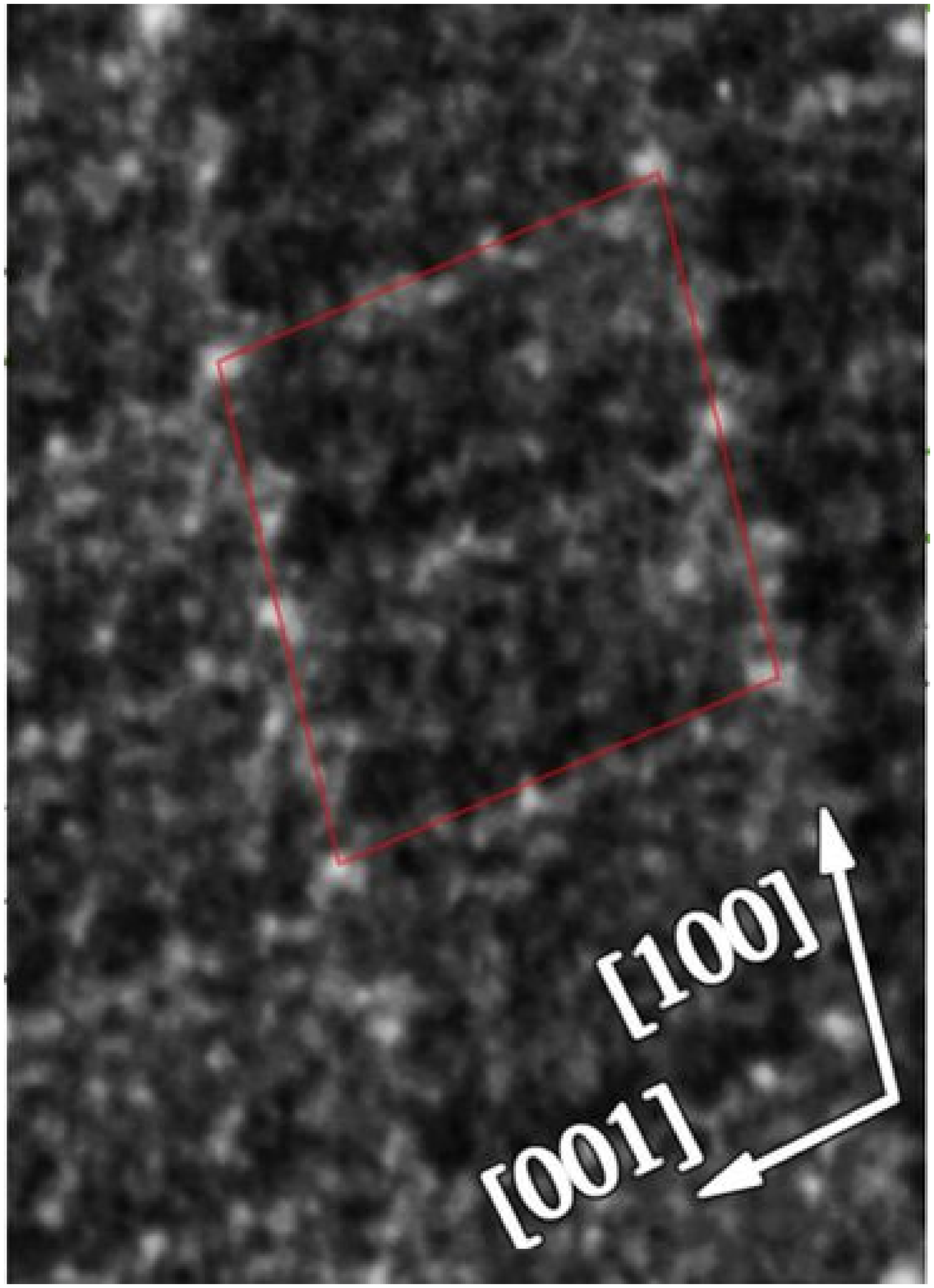}}
\hfill
\subfigure[\label{extended100b}]{\fig[1.8in]{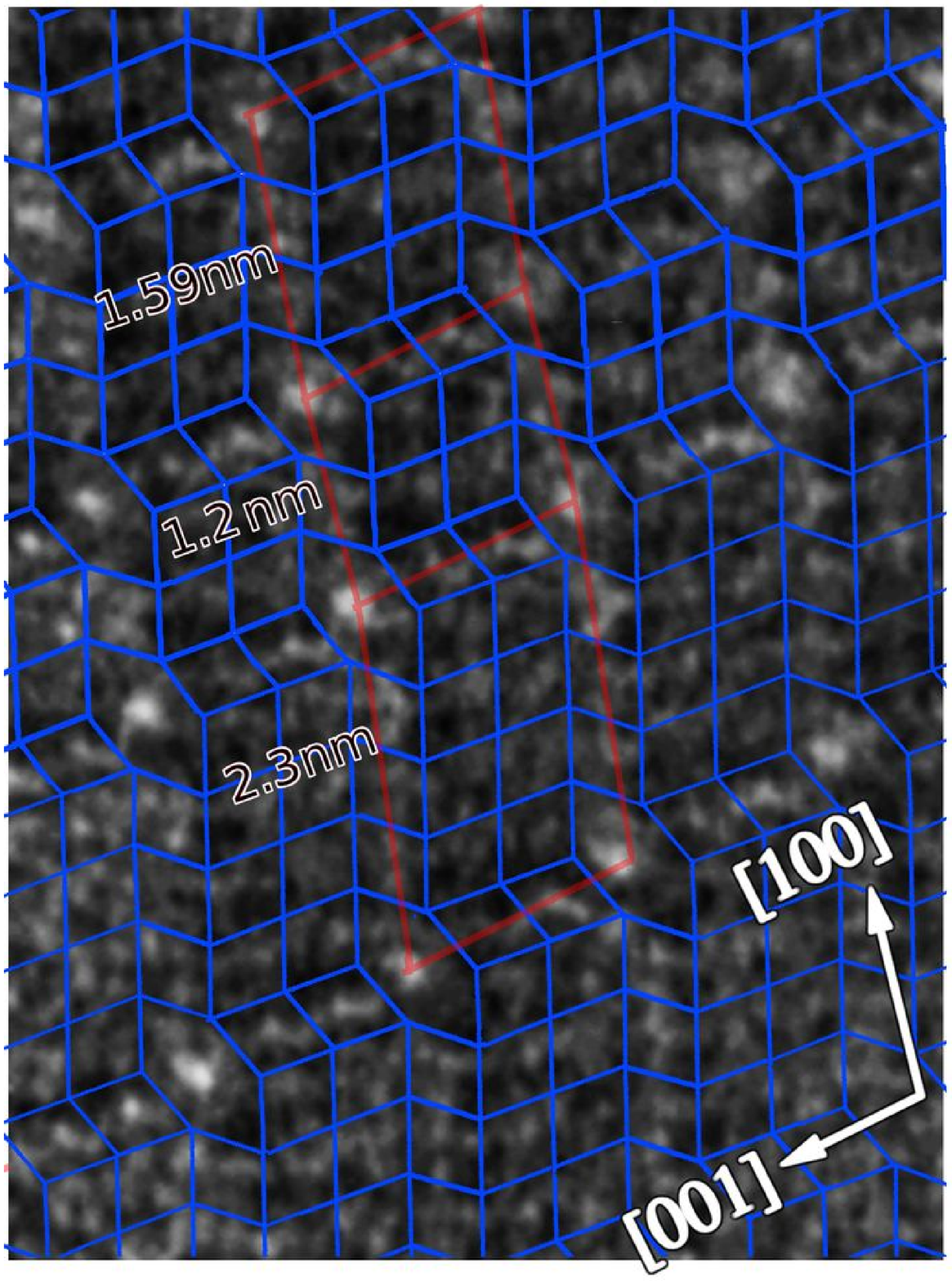}}
\hfill\

\hfill
\subfigure[\label{extended001c}]{\fig[1.57in]{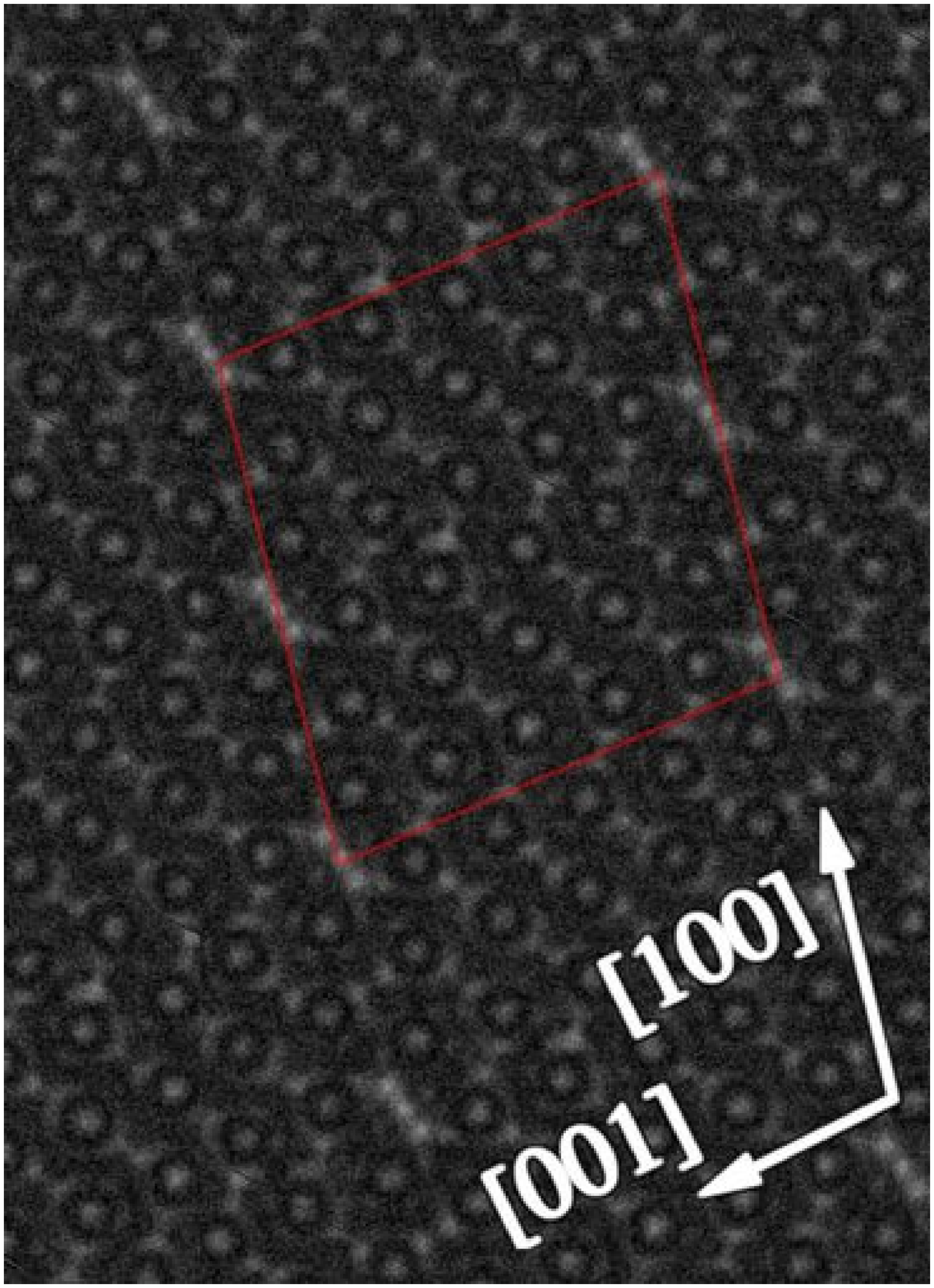}}
\hfill
\subfigure[\label{extended100c}]{\fig[1.8in]{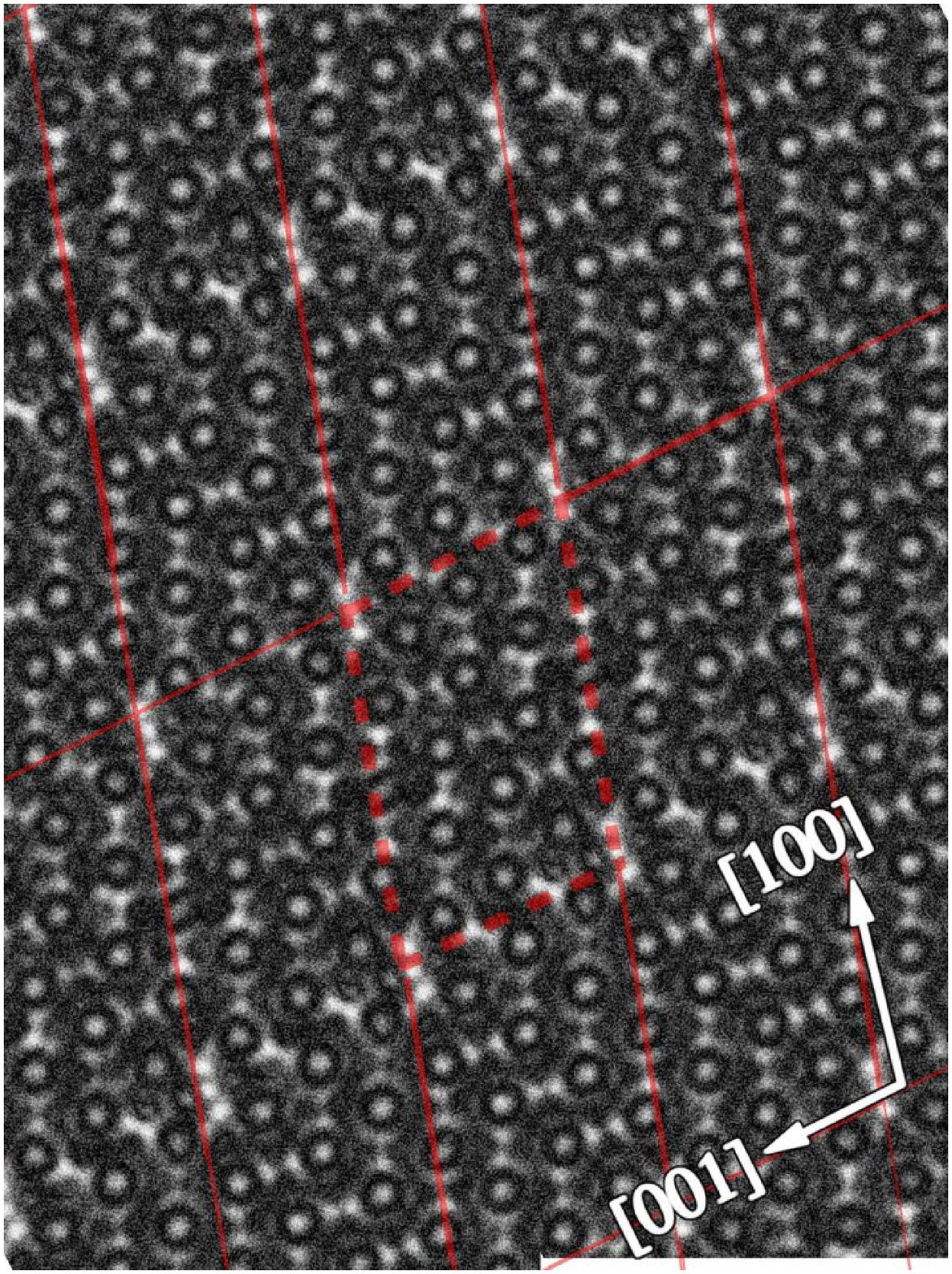}}
\hfill\

\caption{Multi-slice simulations for the proposed transition structures, extended by two acute rhombic units in the [100] direction (a,c,e) and [001] direction (b,d,f). 
a,b) Show the unit cell, atomic sites and the acute rhombic units.
c,d) Show the experimentally obtained images with the unit cells for the enlarged cell (solid lines).
The unit cell for the \ce{Mg4Zn7} structures has superimposed on c).
e,f) Show  multi-slice simulations for thickness 43\,nm at a defocus of 75\,nm. 
Note that the unit cell extended in the [100] direction simulation matches well in the lower portion, whereas the \ce{Mg4Zn7} structure is more accurate for the upper region, suggesting that the transition region has an overall thickness of a half-unit cell of the extended structure. 
\label{extended-sim}}
\end{center}
\end{figure}

Simulations for the transition structures of regions $A$ and $B$  are shown in Figure~\ref{extended-sim}. 
The proposed transition structure for region $A$ is shown in Figure~\ref{extended100a}, with the unit cell 
and rhombic units overlaid. 
This structure was generated from that of \ce{Mg4Zn7}, but with the addition of two additional rhombic units in the [001] direction, resulting in an enlarged monoclinic structure with the same symmetry as the original phase. 
There is good agreement between the experimental image (Figure~\ref{extended100b}) and simulation (Figure~\ref{extended100c}), 
both of which appear very similar to the \ce{Mg4Zn7} structure, except for the greater spacing of the strongly contrasting sites Mg-rich.
The transition region for this interfaces is equivalent in width to one unit cell of the extended structure.

Region  $B$ is somewhat more complex.
The transition region for this (100) interface had a width of 4.1\,nm, but with considerable variation within this region. 
In the experimental image  (Figure~\ref{extended100b}) three regions can be identified, 
based on the spacing of the strongly contrasting sites in the [100] direction. 
The central region has a spacing of $\sim$1.25\,nm between such sites, 
equivalent to $\frac{a}{2}=1.248$\,nm for the \ce{Mg4Zn7} structure. 
Adjacent to this are bands with spacings of 1.59\,nm (above) and 2.3\,nm (below) between such sites.
In the rhombic units overlaid on the image it can be seen that this corresponds to the addition of one and two rhombic units of the \ce{Mg4Zn7} structure per half-unit cell length.
These three regions correspond to half-unit cells ($0<x<\frac{1}{2}a$) of the \ce{Mg4Zn7} structure and two related structures with the addition of more rhombic units between the strongly-contrasting magnesium-rich sites. 
Figure~\ref{extended100a} sets out a structure based on that of \ce{Mg4Zn7} and with the same $C2/m$ symmetry, 
but extended by four rhombic units in the [100] direction, in which a half-unit cell length 
A half unit cell length of this structure (indicated by a dashed line in the simulation in Figure~\ref{extended100c}) matches moderately well with the observed spacing of 2.3\,nm.

The transition structure were also  clearly related to the \ce{Mg4Zn7} and \ce{MgZn2} phases in terms of orientation, with the tiling of the rhombic and hexagonal blocks appearing to running continuously through interface region. 

\begin{table}
\begin{center}
\renewcommand{\arraystretch}{1.6}
\caption{Lattice parameters of monoclinic transition structures formed by adding acute rhombic units to 
the \ce{Mg4Zn7} structure. 
The values in parentheses are those corresponding to the enlarged monoclinic cells shown in Figure~\ref{hrtem-2} and agree well with the expected values for the proposed structures. \label{cells2}.}
\begin{tabular}{lrlrlrr}
 & \multicolumn{5}{c}{Lattice parameters}  & \\ \cline{2-6}
Structure   &  $a$/nm &  & $c$/nm & &   $\beta$/\degree &\\\hline
\ce{Mg4Zn7}     &    2.596  & &   1.448 & & 102.5 &   \\
\parbox[t]{2cm}{$Tr(A)$\\$\mathrm{Mg_{64}Zn_{118}}$}    &    2.596  & (2.6) & 2.389 & (2.42) & 99.4 & (100)\\ 
\parbox[t]{2cm}{$Tr(B)$\\$\mathrm{Mg_{64}Zn_{122}}$}     &    4.371 & (4.5)* &   1.448 & (1.48) & 107.3 & (109) \\ 
\hline
\end{tabular}
\end{center}
{\small *2.25\,nm for the half-unit cell illustrated in Figure~\ref{extended-sim}.}
\end{table}

The transition structure described above are clearly only examples and it would be possible to construct a range of such structures from the rhombic and hexagonal units described in Section~\ref{mono-laves-str}.
The transition structures are larger in unit cell size than \ce{Mg4Zn7} and the hexagonal units comprise a smaller proportion of the structure, which will have a Mg:Zn ratio between the Laves and \ce{Mg4Zn7} monoclinic structures i.e. compositions will be in the range \( \mathrm{MgZn}_{1.75+x},\ 0 < x < 0.25 \). 
This indicates that there may be gradual variations in composition across the \ce{Mg4Zn7}-\ce{MgZn2} interfaces, however this system appears able to accommodate precipitates with compositions between these two values, so long as such structures as constructed principally of the acute rhombic unit present in both \ce{Mg4Zn7} and \ce{MgZn2} phases. 

\subsection{The origin of the mixed-phase precipitates}

It is quite clear that \betap is a transition precipitate. 
The precipitation sequence within the matrix is well known to commence with the precipitation of \betap 
rods in the under-aged condition, followed by the \betapp plates upon overageing \cite{Sturkey1959}. 
The plate-shaped precipitates in the over-aged condition have long been associated with the \ce{MgZn2} structure,
while the rods present at earlier stages of ageing have been identified as either \ce{Mg4Zn7} \cite{Gao2007,Singh2007} or \ce{MgZn2} \cite{Sturkey1959,Takahashi1973,Mendis2009}.
While both phases examined here are metastable intermediates (the equilibrium phase having a composition MgZn and a rhombohedral structure \cite{ClarkPhase1988}) it is  logical to conclude that the \ce{MgZn2} phase is thermodynamically more stable than \ce{Mg4Zn7}, since that latter is only observed in the under-aged or peak-aged condition.

 In the present work, the  precipitates were examined in a condition close to peak age and the predominant precipitate in the matrix (and twins) were \betap rods, which appeared to have a \ce{Mg4Zn7} structure.
The twin boundary precipitate were considerably larger in volume and consisted of either hexagonal \ce{MgZn2} or dual-phase \ce{Mg4Zn7}-\ce{MgZn2}. 
As such the twin-boundary precipitates appear to have progressed further towards equilibrium that those in the matrix, both in terms of the presence of the \ce{MgZn2} phase and in the greater size which has been achieved.

The dual-phase precipitates were comprised of a central region of the monoclinic phase, 
which was always found to adopt one of the previously reported orientation relationships to either the matrix or the twin.
In contrast, the hexagonal phase sometimes formed with an orientation relationship to the matrix 
(e.g. region B in Figure~\ref{hrtem-2}) and sometimes showed no clear lattice matching to the matrix or twin (e.g. region A in Figure~\ref{hrtem-2}).
Despite this difference, both variants possessed an orientation relationship to the monoclinic phase via the rhombic units. 
This, together with the presence of the monoclinic structure at the centre of each mixed-phase precipitate implies that this phase formed prior to the hexagonal Laves phase, with the latter phase growing on the monoclinic phase. 

It is unclear why the nucleation of \ce{Mg4Zn7} occurs first. 
Though it is not the most stable phase by the phase diagram, the stability of small sized precipitates will be predominantly affected by the interfacial energies.
The morphologies will be dictated by the relative interface energies of the precipitate-matrix interfaces. 
Assuming \betap precipitates to be \ce{Mg4Zn7}, its rod-like morphology is dictated by need to minimise 
the area of the high energy 
\( (010)_\ce{Mg4Zn7}-(0001)_\mathrm{Mg} \) 
interface. 
This optimises matching between the basal planes of the magnesium matrix and the \((001)_\ce{Mg4Zn7}\) planes. 
In the plate-like \betapp precipitates, \( (0001)_\ce{MgZn2}-(0001)_\mathrm{Mg} \) interface appears to be a preferred, low energy interface and its area is maximised in this morphology.

The dual-phase precipitates are formed directly in contact with the twin boundary and  it is clearly necessary to recognise that the twin boundary presents a different environment from the matrix.
For example, pipe diffusion along the twin boundary would provide solute at accelerated rates compared to the bulk \cite{Singh2007}, 
assisting in the formation of larger precipitates along the twin interface. 
In addition, both rod shaped and twin boundary precipitates are of nanometer scale, 
with high surface energy to volume ratios, 
and the interfacial energies  may play a significant role in controlling precipitation. 
As the precipitates grow and their interface to volume ratio decreases, the phase equilibria may change. 
A modification of the structure may occur to lower interface energies.
The change in orientation of the Mg phase across the twin boundary is therefore particularly important since the precipitate is not able to adopt the preferred orientation (and hence form the preferred interfaces) to both matrix and twin.

One addition factor that may favour the formation of \ce{MgZn2} in the twin boundary precipitates is the existence of \ce{Mg4Zn7} at the twin interface.
Given the structural similarities in the form of these  rhombic subunits between the \ce{Mg4Zn7} and \ce{MgZn2} phases, and the alignment of such units, (as described in Section~\ref{mono-laves-str}) it is proposed that this feature allows the \ce{Mg4Zn7} phase to provide a favourable surface on which the Laves phase can precipitate.
This kind of structurally-similar interface may have permitted the hexagonal phase to form, even when (as with region A in Figure~\ref{hrtem-2}) it does not possess a good lattice match with the matrix or twin.
This factor would permit the ready nucleation of the \ce{MgZn2} phase at the twin boundaries where pre-existing \ce{Mg4Zn7} precipitates provide suitable nucleation sites.

The co-existence of the monoclinic and hexagonal phases in the twin-boundary precipitates may provide an explanation for the identification of the rod-shaped \betap precipitates as either
\ce{MgZn2} \cite{Sturkey1959,Takahashi1973,Mendis2009} or \ce{Mg4Zn7} \cite{Gao2007,Singh2007}.
It has been shown here that the two phases can co-exist in these larger, twin-boundary precipitates, while maintaining a clear correspondence between the rhombic units within the two structures.
In addition, the presence of several structurally-similar transitional phases at the \ce{Mg4Zn7}-\ce{MgZn2} interfaces indicates that considerable structural and compositional variability can be accommodated within the precipitates. 
A related investigation focused specifically on the rod-like precipitates also identified both \ce{Mg4Zn7} and \ce{MgZn2} within the rod-like precipitates \cite{Singh2009coexist}.
It appears probable therefore, that the rod-like \betap precipitates could be composed of either or both monoclinic and  hexagonal phases plus a wide range of transitional structures, giving rise to the micro-domain structure reported in the literature \cite{Gao2007,Singh2007}.

\subsection{Structural relationship with quasi-crystalline phases}

The \ce{Mg4Zn7}, \ce{MgZn2} and transition structures are essentially modular assemblies or similarly co-ordinated atoms. 
As such, it is worthwhile to consider their relationship, not only to one another but also to the quasi-crystalline i-phase found in Mg-Zn-Y alloys.

Although the \ce{Mg4Zn7} phase has been considered as related to the icosahedral phase directly \cite{Yang1987}, it can be better described as a decagonal quasi-crystal approximant. 
Coexistence of decagonal phase and \ce{Mg4Zn7} phase and the orientation relationship between them has been reported as \(10f\parallel[010] D2a(G)\parallel[-401]\) ; 
and \(D2b(F)\parallel [102]\) \cite{Yi2002}. 
Approximant phases to decagonal quasi-crystals can be orthorhombic or monoclinic. 
Their structure can generally be described as layered. In case of monoclinic approximants, the unique axis corresponding to the decagonal axis, and the monoclinic angle is often close to 72\degree (i.e. $\frac{2\pi}{10}$). 
The length of the unique axis is nearly equal to (or a multiple of) the periodicity of the decagonal phase in that alloy system. 
A well known decagonal approximant in aluminium alloys is $\mathrm{Al_3Fe}$, or $\mathrm{Al1_3Fe_4}$ \cite{Black1956,Henley1985,Kumar1987}.

A decagonal phase was reported in Zn-Mg-RE alloys\cite{Sato1997}. 
The periodicity of this decagonal phase (along the periodic decagonal axis) is reported to be 0.255 nm. 
This is comparable to the lattice spacing along (020) of \ce{Mg4Zn7} and (0002) of magnesium. 
Two crystalline phases related to the decagonal phase were identified to be \ce{MgZn2} and \ce{Mg4Zn7}, the latter of which coexisted with the decagonal phase \cite{Abe1998}. 
Compositionally, only a 2at\% addition of Dy to the \ce{MgZn2} phase formed the composition of the $\mathrm{Zn_{58}Mg_{40}Dy_2}$ decagonal phase\cite{Sato1998}.

In hexagonal phases related to quasi-crystalline phases, the hexagonal axis corresponds to icosahedral twofold axis \cite{Bendersky1987,Singh1998,Kreiner1997}. 
There are icosahedral clusters in three orientations, making the hexagonal symmetry. 
The \ce{MgZn2} phase, however, is a simpler phase in which the hexagonal axis corresponds to a threefold axis of the icosahedral coordination. 
All icosahedra are thus similarly oriented. 
When viewed along a $\langle11\overline{2}0\rangle$ axis, pentagonally coordinated layers are observed, 
as in Figure~\ref{fig-rhomb-laves}.
These pentagons are arranged in a row and have opposite orientations in the layers at two different levels (making two parts of icosahedra in the same orientation).

Often, domains of two phases in a grain are known to create diffraction patterns resembling those from a decagonal phase, such as in Al-Mn\cite{FitzGerald1998,VanTendeloo1991,Li1992} and Al-Mn-Ni \cite{VanTendeloo1988}. 
In the present study, which is on Mg-Zn-Y alloys, the coexistence of the two phases related to decagonal quasi-crystal and their intimately mixing in each other is analogous. 

In the present phases, \ce{Mg4Zn7} and \ce{MgZn2}, 
the rhombic tiles of the intermediate layers form the structural units. 
The vertices of these rhombuses form the vertices of the icosahedral units. 
In the \ce{Mg4Zn7} structure, the main rhombic units, forming a unit of four, have an angle 72.5\degree, 
as shown  at $C$ in Figure~\ref{fig-rhomb-mono}. 
These units are connected by rhombuses with an angle 71.5\degree. 
There is one more type of rhombus with an angle 67.5\degree, and a triangular and a near rectangular unit. 
In between the unit cells, a thin rhombus ($D$ in Figure~\ref{fig-rhomb-mono}) can be defined, with an angle of about 37\degree. 
The \ce{MgZn2} structure is composed of only one type of rhombus with an angle $\sim$72\degree, whose direction alternates ($C^\prime$ in Figure~\ref{fig-rhomb-laves}). 
The two domains of \ce{MgZn2} phase form on either side of the \ce{Mg4Zn7} phase main rhombuses and therefore make an angle of about 72.5\degree with each other. 

\section{Conclusions}

Precipitation on twin boundaries has been studied in a quasi-crystalline-bearing Mg-Zn-Y alloy.
Twins were introduced in the alloy by extrusion followed by a 2\% deformation of the extruded rod in compression. 
The samples were solutionised before the compressive deformation. On isothermal ageing of the deformed samples at 150$^o$C, 
several forms of precipitation occurred. 
Well known thin rod-like \betap precipitates formed in the matrix (and twins) along the $[0001]_\mathrm{Mg}$ axis. 
Bulkier precipitates formed at the twin boundaries. 
These precipitates were either of the hexagonal phase with the orientation relationship 
\( [0001]_\mathrm{Mg} \parallel[\overline{1}2\overline{1}0]_\ce{MgZn2}\); 
\( \{11\overline{2}0\}_\mathrm{Mg}\parallel(0001)_\ce{MgZn2}\) with either the matrix or the twin, 
or had a core of monoclinic phase \ce{Mg4Zn7} with \ce{MgZn2} phase growing in two orientations, $A$ and $B$ over it. 
The \ce{Mg4Zn7} phase was oriented with OR 
\([0001]_\mathrm{Mg} \parallel[010]_\ce{Mg4Zn7}\); 
\( \{\overline{1}010\}_\mathrm{Mg}\parallel(201)_\ce{Mg4Zn7}\) with either the matrix or the twin. 
Both the orientations of the \ce{MgZn2} phase growing over the \ce{Mg4Zn7} phase had axial orientation 
\([010]_\ce{Mg4Zn7}\parallel[\overline{1}2\overline{1}0]_\ce{MgZn2}\). 
In orientation $A$, \((20\overline{1})_\ce{Mg4Zn7} \sim (10\overline{1}2)_\ce{MgZn2}\) and in orientation $B$ \((20\overline{2})_\ce{Mg4Zn7} \sim (10\overline{1}\overline{2})_\ce{MgZn2}\). 
For \ce{MgZn2} phase in orientation $B$, the OR with the matrix was 
\([0001]_\mathrm{Mg}\parallel[\overline{1}2\overline{1}0]_\ce{MgZn2}\);
\(\{11\overline{2}0\}_\mathrm{Mg}\parallel(0001)_\ce{MgZn2}\). 
\ce{MgZn2} phase orientation $A$ formed with 
\([0001]_\mathrm{Mg} \parallel [\overline{1}2\overline{1}0]_\ce{MgZn2}\); 
\( \{\overline{1}010\}_\mathrm{Mg} \sim (\overline{1}012)_\ce{MgZn2}\) with the matrix. 
Both the phases \ce{Mg4Zn7} and \ce{MgZn2} are modular structures composed of similar units of icosahedrally coordinated atoms. In the layers comprised of zinc atoms, which make vertices of the icosahedral units, "thick" rhombic tiles can be drawn. 
Rearrangement of these tiles brings about the change in the structure. 
An orientation relationship was identified between these rhombic tiles which retained a common alignment across the \ce{Mg4Zn7}-\ce{MgZn2} interface even in the $A$ orientation which lacked an exact lattice match with the matrix or monoclinic phase.
Addition of these tiles to the \ce{Mg4Zn7} structure makes the transition structures, increasing either the a or the c parameters. 
Two transition structures are identified here, whose unit cells are (A) a=2.596 nm, b=0.524, c= 2.389 and =99.4$^o$, and (B) a=4.371, b=0.524, c=1.448 and =107.3$^o$. 
These transition structures have stoichiometry between the \ce{Mg4Zn7} and \ce{MgZn2} phases. Similarities to quasicrystal-related complex phases in aluminium-based alloys have been pointed out.

\section*{Acknowledgements}

One of the authors (JMR) gratefully acknowledges the support of the Japan Society for the Promotion of Science (JSPS) through a JSPS fellowship. 
The authors also thank  Reiko Komatsu, Keiko Sugimoto and Toshiyuki Murao for assistance with sample preparation.


\enlargethispage{1cm}

\end{document}